\documentclass[aps,preprintnumbers,amsmath,amssymb,prd,superscriptaddress,longbibliography,twocolumn]{revtex4-1}
\usepackage{color}
\usepackage{subfigure}
\usepackage{feynmp} 
\usepackage{feynmp-auto}
\usepackage{pgfplots}
\usepackage[breaklinks=true,unicode=true,urlcolor = blue,colorlinks = true,citecolor = blue,linkcolor = blue]{hyperref}

\usepackage{graphicx}
\usepackage{dcolumn}
\usepackage{stix}
\usepackage{hyperref}
\usepackage[version=4]{mhchem}
\usepackage{array}
\usepackage{booktabs}
\newcommand{\myref}[1]{Eq. (\ref{#1})}
\counterwithout{figure}{section}

\usepackage{float}
\usepackage[marginal]{footmisc}
\usepackage{color}
\usepackage{url}

\begin{document}
	\title{Deviations from Arrhenius dynamics in high temperature liquids, a possible collapse, and a viscosity bound}	
	\author{Jing Xue}
	\affiliation{Department of Electrical and Computer Engineering, Lehigh University, Bethlehem, PA 18015, USA}
	\author{Flavio S. Nogueira}
	\affiliation{Institute for Theoretical Solid State Physics, IFW Dresden, Helmholtzstr. 20, 01069 Dresden, Germany}
	\author{K. F. Kelton}
\affiliation{Institute of Materials Science and Engineering, Washington University, St. Louis, MO 63130, USA}
	\affiliation{Department of Physics, Washington University, St.
Louis, MO 63160, USA}
	\author{Zohar Nussinov}
	\affiliation{Department of Physics, Washington University, St.
Louis, MO 63160, USA}
	\date{\today}

\begin{abstract}
Liquids realize a highly complex state of matter in which strong competing kinetic and interaction effects come to life. As such, liquids are, generally, more challenging to understand than either gases or solids. In weakly interacting gases, the kinetic effects dominate. By contrast, low temperature solids typically feature far smaller fluctuations about their ground state. Notwithstanding their complexity, with the exception of quantum fluids (e.g., superfluid Helium) and supercooled liquids (including glasses), various aspects of common liquid dynamics such as their dynamic viscosity are often assumed to be given by rather simple, Arrhenius-type, activated forms with nearly constant (i.e., temperature independent) energy barriers. In this work, we analyze experimentally measured viscosities of numerous liquids far above their equilibrium melting temperature to see how well this assumption fares. We find, for the investigated liquids, marked deviations from simple activated dynamics. Even far above their equilibrium melting temperatures, as the temperature drops, the viscosity of these liquids increases more strongly than predicted by activated dynamics dominated by a single uniform energy barrier. For metallic fluids, the scale of the prefactors of the best Arrhenius fits for the viscosity is typically consistent with that given by the product ($nh$) with $n$ the number density and $h$ Planck's constant. More generally, in various fluids (whether metallic or non-metallic) that we examined, $(nh)$ constitutes a lower bound scale on the viscosity. We find that a scaling of the temperature axis (complementing that of the viscosity) leads to a partial collapse of the temperature dependent viscosities of different fluids; such a scaling allows for a functional dependence of the viscosity on temperature that includes yet is far more general than activated Arrhenius form alone. We speculate on relations between non-Arrhenius dynamics and thermodynamic observables. 
 \end{abstract}
\maketitle

\section{Introduction}
The Arrhenius equation \cite{Arrhenius1889p96,Arrhenius1889p226,Arr2, Arr1, Petrowsky2013} is an empirical relation describing the relationship between the reaction rate and the temperature $T$ of a chemical reaction \cite{reactionrate55yr,reactrate}. The reaction rate constant $k(T)$ quantifies the speed at which the reaction occurs. The Arrhenius equation asserts that
\begin{equation}
\label{arr:eq}
k(T) \propto e^{-E_a/k_{B}T}.
\end{equation}
Here, $E_a$ is an ``activation energy'', and $k_B$ is the Boltzmann constant. 
Another expression, commonly derived in transition rate theory textbooks, the Eyring equation \cite{eyring1}, contends that the transition rate is, more precisely, given by
\begin{equation}
\label{eyring:eq}
k(T)=\frac{\kappa k_{B}T}{h}e^{-\Delta G/k_{B}T}.
\end{equation}
In \myref{eyring:eq}, the constant $\kappa$ is the ``transition coefficient'' and $h$ is Planck's constant. Similar to the Arrhenius equation (\myref{arr:eq}), the reaction rate in \myref{eyring:eq} depends exponentially on the Gibbs free energy of activation $\Delta G$- which assumes the role of a barrier. This free energy barrier $\Delta G=\Delta H-T\Delta S$ generally includes both entropic ($\Delta S$) and enthalpic ($\Delta H$) contributions \cite{gibbs1}. A weakly temperature dependent $\Delta G$ qualitatively emulates a dominant exponential decay (\myref{arr:eq}) with a constant $E_a$. In the current work, we will synonymously use $E_a$ and $\Delta G$ to denote the effective (free) energy activation barriers. Related transition state forms have been posited over the years \cite{reactionrate55yr, reactrate}. 

Beyond its historical roots in chemical reaction rates, the Arrhenius equation has seen widespread use in other (at times, interrelated) arenas including (i) semiconductor physics (e.g., where it enables a determination of the number of thermally activated electrons in the conduction and valence bands) \cite{AM} aiding theoretical design and enabling a basic understanding of diodes, transistors, solar cells, and other semiconductor devices, (ii) metallurgy (e.g., creep rate and the number of vacancies/interstitial sites in a crystal), e.g., \cite{creep1,creep2,creep3,creep4}, (iii) the analysis of data from dynamical probes such as those of dielectric response, NMR and NQR in a host of systems, e.g., \cite{Cava,Curro,NMR,NMR1,NMR2}, (iv) relaxation rates associated with particles of fixed structural ``softness'' (an analogue of elastic defect density in amorphous systems whose average value correlates with the viscosity) \cite{softnessliu} and, notably, (v) fluid dynamics- the focus of our work. 

Before proceeding further, we must briefly comment on a well known exception to activated liquid dynamics- that of supercooled fluids, e.g., \cite{Langer2014,supercool1996,supercool1,supercool2,supercool3,supercool4,Angell1924,relaxAngell,qmrelaxtime1,kelton2016,correlation2016,relaxArr2012, berthier2011, BerthierBiroli2011}. Compounding the silicates that have been known to form glasses since antiquity, numerous liquids may be experimentally supercooled below their ``freezing'' or liquidus temperature $T_{l}$. Such a rapid cooling does not enable the liquids to change their phase and thus crystalize at the equilibrium freezing temperature. At low enough temperatures, these supercooled liquids assume a glassy amorphous state. During the supercooling process, the typical relaxation time scale of the liquids may increase dramatically by many orders of magnitude for a modest temperature drop and strongly deviate from the Arrhenius form of \myref{arr:eq}. This capricious divergence from activated dynamics is one of the principle features underscoring the enigmatic character of supercooled liquids and glasses.  

Excusing the above celebrated exception of supercooled liquids as well as that of quantum fluids at cryogenic temperatures (most notably, low temperature Helium at ambient pressure), the viscosity $\eta$ of most equilibrated liquids (including glassformers at sufficiently high temperatures above their liquidus temperature) has, for decades, been largely presumed to be given by an Arrhenius type expression. Specifically, it is commonly assumed that the typical relaxation time $\tau$ of liquids is given by 
\begin{equation}
\label{eq:0}
\tau=\tau_0 e^{E/T}.
\end{equation}
In \myref{eq:0}, the activation barrier $E_a$ has been rescaled by the Boltzmann constant $k_{B}$ so that it is measured in units of temperature (Kelvin) (i.e., $E \equiv E_{a}/k_{B}$). While in harmonic solids, there is a linear (Hooke law like) relation between stress and strain, in fluids, it is the strain rate (the time derivative of the strain (or displacement)) that is linearly proportional to the stress (or force); this is a continuum counterpart to the linear relation between the applied force and particle velocity in a viscous system (with the ratio between the two  being set by a viscosity). By the Maxwell model for viscoelasticity, the dynamic viscosity scales as
\begin{equation}
\label{eq:99}
\eta={\sf G}\tau,
\end{equation}
where ${\sf G}$ is the instantaneous shear modulus \cite{relaxAngell} and $\tau$ is the relaxation time (inverse strain relaxation rate). From these and other considerations,
viscosities are often anticipated to display a behavior similar to that of \myref{eq:0}, 
\begin{equation}
\label{eq:1}
\eta(T)=\eta_0 e^{E/T}.
\end{equation}
Indeed, e.g., aside from \myref{eq:0}, also the Eyring form of \myref{eyring:eq} leads, in certain treatments \cite{eyring1,qmrelaxtime1} to \myref{eq:1} with a viscosity prefactor $\eta_0$ that scales with the particle number density $n$ (a quantity that is, typically, weakly temperature dependent). Thus, on the whole, the time scale governing liquid dynamics and respective viscosities are often assumed to be effectively governed by a single activation barrier of uniform height that is set by $E$ and with a constant prefactor $\eta_0$. \myref{eq:1} was first noted in a work by Reynolds \cite{Reynolds1886} (three years before Arrhenius introduced his now famous equation for reaction rates). Nonetheless, viscosity satisfying \myref{eq:1} is commonly said to be of the ``Arrhenius'' type due to the intuitive connection, briefly reviewed above, that the Arrhenius rate equation of \myref{arr:eq} evokes. In the many years since, this relation has been posited and rediscovered anew by several researchers (perhaps most notably, by Guzman \cite{Guzman} and Andrade \cite{Andrade} (in some circles, \myref{eq:1} is known as the Guzman-Andrade or Andrade equation)). 

In the current work, we will extensively analyze the temperature dependence of the dynamic viscosities of numerous fluids. Empirically, in accord with certain theoretical anticipations (related to those underlying \myref{eyring:eq}), when fitting the viscosities of various fluids to the Arrhenius form of
\myref{arr:eq}, the prefactor in \myref{eq:1} ($\eta_0$) was, in certain instances, found to be of the scale of a particularly simple product: $(nh)$. Here, $h$ is Planck's constant and $n$ denotes the number of particles per unit volume \cite{eyring1,qmrelaxtime1,Tcoop1,kelton2016,otp1967}. In a more general vein, the product $(nh)$ has thus been suggested to be a lower bound on the scale of the viscosity \cite{qmrelaxtime1}. (A related tighter bound differing by factors of mass ratios was later proposed in \cite{newwater}.) In the current work, we will demonstrate that empirically,
\begin{eqnarray}
\label{lowernh}
\eta \ge {\cal{O}}(nh).
\end{eqnarray}
Related lower bounds on the viscosity were further rigorously proven in \cite{hbound}. 

Low viscosity is associated with a high Reynolds number regime where the system may become most turbulent. In conventional materials, the viscosity is minimal at a crossover between the gaseous and fluid viscosity behaviors. In the gas (due to increased collisions between the molecules as the gas is heated and thus an effective increase in the coupling between layers), the viscosity increases with temperature. By contrast, in the fluid, the viscosity monotonically drops with temperature due to increased thermal motion which effectively reduces the coupling between fluid layers (the interactions become less important relative to thermal effects). Fluids with high viscosity support laminar flow wherein shear stresses readily dampen applied perturbations. The two opposing behavioral trends of monotonic decrease and increase in the viscosity as a function of temperature in, respectively, the fluid and the gas mandate an experimentally observed intervening viscosity minimum (that, as noted above, empirically satisfies \myref{lowernh}).

Broader than the single uniform activation barrier of \myref{eq:1}, any function $\eta(T)$ may be written as a Laplace transform in the inverse temperature $\beta = \frac{1}{T}$ (after, once again, rescaling the activation energies by $k_{B}$) via a distribution $P(E')$ of effective energy barriers,
\begin{eqnarray}
\label{Laplace1}
\frac{1}{\eta(T)} = \frac{1}{\eta_{0}} \int P(E') e^{-\frac{E'}{T}} dE'.
\end{eqnarray}
As \myref{Laplace1} emphasizes, in principle, an inverse Laplace transform of any data set over an extensive range of temperatures will trivially yield an activation energy distribution $P$ that will, by construction, reproduce the measured viscosity data. Physically, the distribution $P$ may be not only a function of the activation energy alone but also of the temperature. In a similar spirit, for the viscosity data of liquids below their liquidus temperature $T_{l}$ (not above it as we focus on in the current work), a particular theory\cite{supercool1,supercool2,supercool3,supercool4} reproduces all experimentally measured data of supercooled liquids over 16 decades of viscosity with a single-parameter scale free temperature dependent normal distribution $P$ of effective equilibrium relaxation rates. In what follows, we will test the applicability of \myref{Laplace1} with a delta function distribution $P(E') = \delta(E-E')$ associated with the commonly assumed activation energy form of \myref{eq:1} for high temperature liquids (above their melting temperature). Throughout this work, {\it all values of the dynamic viscosity $\eta$ (as well as the scale $\eta_{0}$) will be quoted in units of $Pa \cdot s$} (= 0.1 Poise). In order to provide an everyday intuitive feel for these conventional units, we remark that the viscosity of water at room temperature is $\sim$ 1centiPoise $=$ milli-$Pa \cdot s$. While numerous investigations applied the fit of \myref{eq:1} for the dynamic viscosities to various fluids, we are not aware of much work that critically focused on and tested the veracity of the Arrhenius form to these viscosities and any deviations thereof. It is important to underscore that notwithstanding their simple intuitive appeal, there are no rigorous derivations of activated dynamics in fluids. Indeed, theoretically determining the viscosity of fluids as a function of their temperature is not, at all, an easy problem. Unlike the solid (where interaction effects dominate) or the gas (where kinetic effects are important), in the fluid, both kinetic effects and interactions are comparable and compete with one another. In this paper, we will test the validity of Arrhenius form of equilibrium fluids (at temperatures $T>T_l$). Towards this end, we will compute, within various temperature intervals, values of effective activation energy $E$ and prefactor $\eta_0$ that match the experimental values of the viscosity when \myref{eq:1} is assumed. These tests will enable us to comment on the high temperature limits of various well known fits to the viscosity of glass forming liquids, e.g., \cite{kkznt1,kkznt2,kkznt3,deh1,Tcoop1}. Various notable works advanced and tested various (kinematic) viscosity fits, e.g., \cite{Seeton}. To our knowledge, there are no prior investigations explicitly focusing on the effective activation energies associated with high temperature liquids. 

In the current work we will frequently refer to {\it two different temperatures}:

(1) In several earlier studies of glass forming fluids, e.g., the above noted \cite{kkznt1,kkznt2,kkznt3,Tcoop1}, {\it a crossover temperature} $T_A$ was identified below which ($T<T_A$) strong deviations from Arrhenius behavior were found and above which (temperatures $T>T_A$) the dynamics seemed to conform to an approximate Arrhenius form.

As we will detail, our analysis reveals that, as a general trend, even up to temperatures far above melting, disparate fluids may exhibit a viscosity that varies more significantly with temperature than activated dynamics (\myref{eq:1}) would predict. In other words, the commonly assumed simple Arrhenius form does not, in fact, accurately capture fluid dynamics. Towards that end, 

(2) We tested for {\it a broader temperature dependence} associated with {\it a general scaling temperature $T_{sc}$}. Specifically, we examined whether scaled dimensionless viscosities of different fluids $\eta/\eta_{0}$
can, in their high temperature regime, be made to collapse as a function of corresponding dimensionless inverse temperatures $T_{sc}/T$. Here, $\eta_{0}$ and $T_{sc}$ are fluid specific parameters. The Arrhenius form of \myref{eq:1} is only one realization of such a possible collapse; in the Arrhenius form, the activation energy $E$ is set by $T_{sc}$ and the function specifying the associated collapse is the exponential function. We searched for such a more general possible collapse and determined the optimal associated temperature scales $T_{sc}$ over a broad range of viscosities. Our tests revealed that certain fluids exhibit nearly identical dependences on the temperature over 13 decades of viscosity while others collapse over a far more limited range.

\section{Outline}

The remainder of this article is organized as follows. We begin, in Section \ref{linear_invariance:sec}, 
by describing simple litmus tests indicating deviations from Arrhenius dynamics with constant activation energies. 
We then turn (Section \ref{sec:test}) to briefly discuss the various fluids that we analyzed and the smoothening procedure invoked in our numerically evaluated derivatives. Sections \ref{sec:activation_barriers},\ref{sec:eta0}, and \ref{effEnt} quantify, via complementary calculations, the deviations from simple Arrhenius dynamics. Taken together, these analyses illustrate how the effective activation barriers typically increase as the temperature is lowered. In Section \ref{collapse:sec},
we demonstrate that it is possible to collapse the viscosity data from different fluids onto a curve with some liquids persisting for many decades of the viscosity for while others such a collapse is over a very limited range. In Section \ref{sec:comparelnnh}, we illustrate that the lower bound \cite{qmrelaxtime1} on the viscosity of \myref{lowernh} holds for all materials investigated and contrast it with other more recent proposed bounds \cite{newwater}. We conclude with a synopsis of our results in Section \ref{sec:conclusions}. Various technical details have been relegated to the Appendices.

\section{Hallmarks of deviations from simple activated dynamics in equilibrated high temperature fluids}
\label{linear_invariance:sec}

Since the viscosity of various fluids typically span several decades as their temperature is varied, we will analyze \myref{eq:1} on a logarithmic scale,
\begin{equation}
\label{eq:2}
\ln\eta(T)=\ln\eta_0+\frac{E}{T}.
\end{equation}

We now explicitly highlight the exceedingly simple principles underlying much of our study. 
If \myref{eq:2} holds, then both $E$ and $\eta_{0}$ as adduced from (numerical) derivatives of the experimentally measured viscosities of all studied liquids should be temperature independent constants. That is, for simple activated dynamics, both
\begin{eqnarray}
\label{eq:4}
E=\frac{d}{d(1/T)}\ln\eta,
\end{eqnarray}
and 
\begin{equation}
\ln\eta_{0}=\frac{d }{d T}(T\ln\eta)
\label{eq:200}
\end{equation}
must, for each individual fluid, assume the same value at all temperatures. Temperature variations of the derivatives of Eqs. (\ref{eq:4}, \ref{eq:200}), if any exist, will attest to the degree to which departures from the putative Arrhenius form of \myref{eq:1} appear.

\begin{figure}[h]
	\centering
	\includegraphics[scale=1]{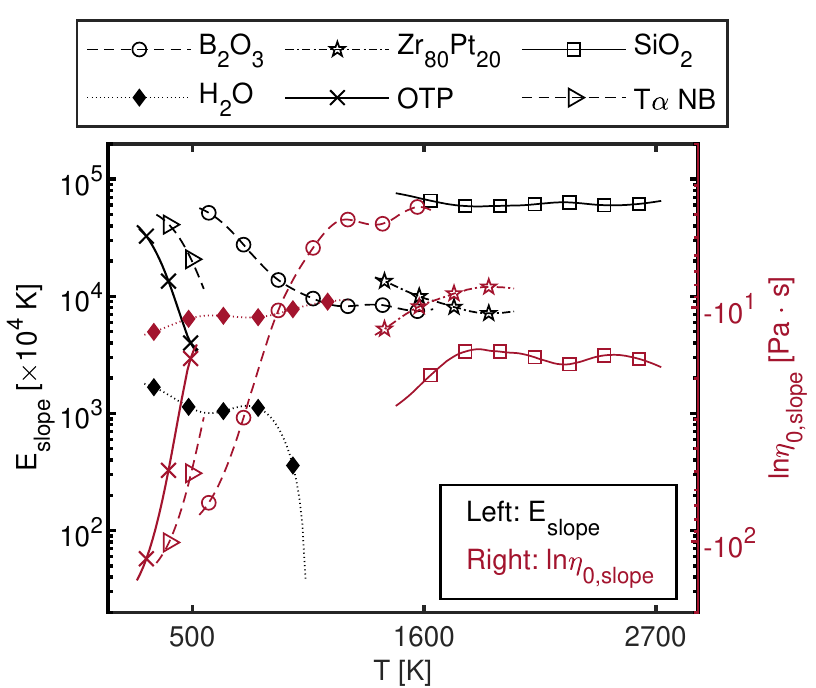}
	\caption{(Color online.)	Finite difference numerical (``slope'') approximations of the effective activation energy $E$ of \myref{eq:4} (as determined by the finite gradient approximation of Eq. (\ref{eq:60}), see left vertical axis with points marked in black) and of $\eta_0$ of \myref{eq:200} (as computed via the finite difference approximation of Eq. (\ref{eq:201}), see right vertical axis with points highlighted in red) for six liquids above their liquidus temperature $T_l$. Notwithstanding fluctuations, it is seen that across all fluids, as the temperature decreases, the finite difference approximation to the effective activation energy $E(T)$ becomes larger. The ``strong'' glass forming fluid $\ce{SiO2}$ \cite{vollmayrSiO2} exhibits a nearly constant large effective energy barrier at the high temperature shown (far above that of supercooling and glass formation). The finite difference approximations to $\eta_0$ as computed from \myref{eq:200} increase as temperature rises. Combined with their more notable deviations from Arrhenius dynamic, the sparse viscosity data of $\ce{B2O3}$, OTP and T$\alpha$NB give rise to curves that are not very smooth.}
	\label{fig: eeta0part}
\end{figure}

Given the above corollaries for temperature independent $E$ and $\eta_{0}$ as determined from derivatives, we display in Fig. \ref{fig: eeta0part} (see left vertical axis) our approximate numerical for the activation energy of \myref{eq:4} as deduced from experimentally measured viscosities for all liquids which we analyzed above their liquidus (or melting) temperature $T_{l}$. Similarly, on the right vertical axis of Fig. \ref{fig: eeta0part}, we provide our numerically evaluated results for \myref{eq:200} at temperatures about the equilibrium liquidus temperature.

As Fig. \ref{fig: eeta0part} makes evident, empirical numerical approximations to the derivatives of Eqs. (\ref{eq:4}, \ref{eq:200}) are not temperature independent constants. Specifically, as Fig. \ref{fig: eeta0part} attests for temperatures $T>T_{l}$ (and, respectively, Figs. \ref{fig:e_eta0_s_metallic}, \ref{fig:e_eta0_s} of Appendix \ref{ReliabilityofResults} demonstrates for $T>T_{sc}$), at the lower end of the temperature range (and usually at far higher temperatures as well):  \newline

{\it The effective activation energy as computed from \myref{eq:4} increases significantly as the temperature $T$ decreases.}  \newline

In order to quantitively contrast the viscosity with the Ansatz of \myref{eq:1} with empirical results, we may endow $\eta_{0} \to \eta_{0}(T)$ and/or $E \to E(T)$ with temperature dependences so as to optimally fit the experimental viscosity data. (Varying any one of these parameters alone will suffice to fit any experimental data set.) By fiat, the activated form of \myref{eq:1} assumes the absence of any such temperature dependences. If \myref{eq:1} applies, with moderate variations of $E(T)$ and $\eta_{0}(T)$, then (since $\eta$ is exponential in $E$) a numerical evaluation of \myref{eq:200} will capture the derivative $dE/dT$ of the effective activation barrier. The marked decrease of the numerically evaluated \myref{eq:200} in Fig. \ref{fig: eeta0part} indicates that the rate of increase of the effective activation barrier $E(T)$ as the temperature $T$ is decreased becomes more pronounced as the temperature is lowered. Similar trends (made evident in Fig. \ref{fig:e_eta0_s}) are also visible at temperatures above the (typically) higher temperature crossover temperature $T_A$ \cite{kkznt1,kkznt2,kkznt3,deh1,Tcoop1}. The astute reader may note from these figures that for water, at sufficiently high temperatures, some of these trends are reversed. This is so since the extended temperature range that we investigate also includes temperatures above the boiling point of water (373 K) where the system is no longer a liquid. Indeed, while an increase in temperature typically decreases the viscosity of the liquid, in a gas this trend is reversed.

\myref{eq:2} is trivially invariant under the simultaneous transformations
\begin{eqnarray}
\label{gaugef}
\eta_{0}(T) \to  \eta_{0}(T) e^{- \frac{f(T)}{T}}, \nonumber
\\ E(T) \to E(T) + f(T),
\end{eqnarray}
with $f(T)$ an arbitrary function of the temperature. As further discussed in Appendix \ref{estimateEeta}, possible effective temperature dependencies of $E$ and $\eta_0$ may be obtained by, e.g., explicitly plotting
\begin{equation}
\label{eq:3}
T\ln\eta=T\ln\eta_0+E
\end{equation}
as a function of the temperature. The equality of \myref{eq:3} was implicitly invoked in deriving \myref{eq:200} under the assumption of constant $E$ and $\eta_0$.  As seen from \myref{eq:3}, if the ansatz of \myref{eq:1} applies, then a plot of $(T \ln \eta)$ as a function of $T$ will yield a line with a slope set by $\ln \eta_0$ and intercept equal to $E$. Setting, in \myref{gaugef}, $f=aT+b$ (with general constants $a$ and $b$) to be an arbitrary linear function will yield other consistent parameters for any such nearly constant $E$ and $\eta_0$. Taken together, Eqs. (\ref{eq:4}, \ref{eq:200}, \ref{eq:3}) (all which trivially stem from \myref{eq:2}) allow for an estimate of the typical values of $E$ and $\eta_{0}$.
This will also allow us to monitor for any temperature variations from assumed constant values of these parameters. Detailed analyses of various high temperature fluids all lead to our earlier highlighted conclusion: if the viscosity of liquids is fitted to a single uniform activation energy form then the resulting $E(T)$ exhibits, on the whole, an increase as $T$ is lowered. 

In what follows, we briefly discuss the liquids that we examined and our discrete temperature difference approximations to the derivatives of Eqs. (\ref{eq:4}, \ref{eq:200}).

\section{The studied liquids and general aspects of the data analysis}
\label{sec:test}
\begin{figure}[h]
	\centering
	\includegraphics[scale=1]{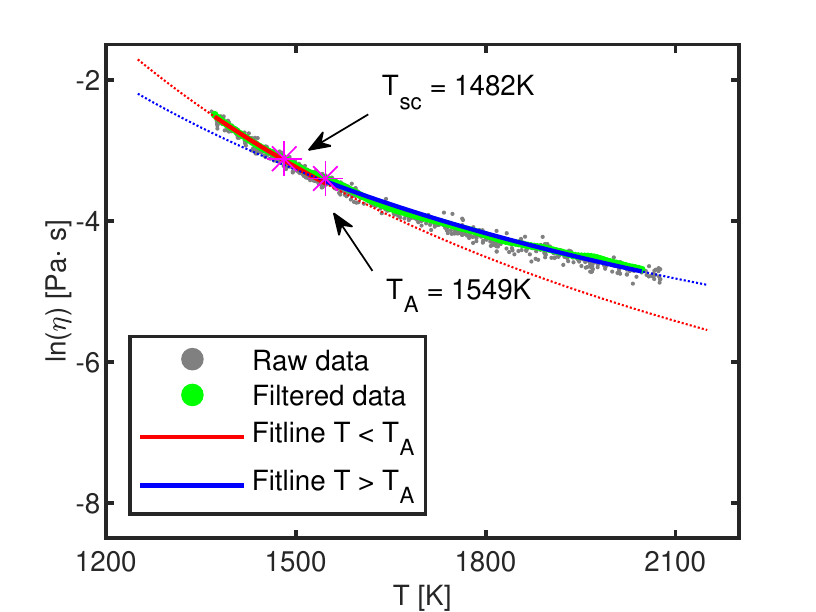}
	\caption{(Color online.) The determination of a crossover temperature $T_{A}$  \cite{kkznt1,kkznt2,kkznt3,deh1,Tcoop1}. A comparison of the temperature-dependent viscosity data for $\rm Zr_{80}Pt_{20}$ and the Arrhenius form of \myref{eq:1}. The gray dots represent the raw experimental data; the green curve marks the ``filtered'' data (i.e., the data averaged over a fixed temperature window to minimize the experimental noise). The solid red line is a fit to \myref{eq:2} of the filtered data for temperatures $T > T_{A}$ (see \cite{Tcoop1,deh1} for the definition and discussion of $T_A$ for general fluids). The dashed red line is an extrapolation of this constant activation energy Arrhenius fit to lower $T$. At higher temperatures, the deviations from the expected constant activation energy of \myref{eq:1} are less pronounced than those at lower temperatures. An approximate crossover appears from Arrhenius behavior at temperatures above $T_A$ to a strongly non-Arrhenius one at lower temperatures. $T_{sc}$ is the scaling temperature (see main text).} 
	\label{fig:zrptraw0}
\end{figure}

We list all of the liquids that we study along with some of their properties (such as the scaling temperature $T_{sc}$ which we describe next) in Table \ref{tab:11}. Our focus is on high temperature behavior (i.e., at temperatures far above the liquidus temperature) where Arrhenius behavior was assumed to universally hold for all liquids. The liquids in Table \ref{tab:11} include good glass forming liquids known to display non-Arrhenius dynamics when supercooled to temperatures lower than $T_l$. For the metallic fluid glassformers \cite{TAphen}, plots of $\ln(\eta/\eta_0)$ as a function of $T_{sc}$, with a material dependent $T_{sc}$ for each individual fluid, collapse onto a single universal curve \cite{Tcoop1}. Fig. \ref{fig:zrptraw0} shows both the raw and filtered data of $\rm Zr_{80}Pt_{20}$ and the corresponding fit of \myref{eq:1}. By ``filtered data'', we refer to the replacement of $\ln \eta(T)$ by the average of $\ln \eta(T)$ over a finite temperature window $[T-\frac{\Delta T}{2}, T + \frac{\Delta T}{2}]$ centered about each temperature $T$. Averages over finite width ($\Delta T$) temperature windows suppress oscillations of the raw empirical values of $\ln \eta(T)$. In Appendix \ref{lowpassfilter}, we further discuss these averages. The linear fit in 
Fig.\ref{fig:zrptraw0} illustrates how a crossover temperature $T_{A}$ (and a scaling temperature $T_{sc}$ for which such a collapse occurs) may be ascertained. 
\footnote{As we will detail in Section \ref{collapse:sec} (and Figure \ref{fig:determineta} therein), a scaling of the temperature by a liquid dependent $T_{sc}$ allows for a collapse  (\myref{Fcoll}) of the viscosity data over a broader range of temperatures than simple Arrhenius dynamics. This is so since the function $F$ in \myref{Fcoll} may differ from the simple exponential of \myref{eq:1}.}

The values of $\ln\eta_0$ listed in Table \ref{tab:11} are those associated with the fit of \myref{eq:1} for temperatures $T<T_{sc}$. The viscosity data of the metallic liquids were measured by one of us \cite{kelton2016}. Other viscosity data were extracted by scanning graphs from published works \cite{water,B2O3,OTPdata,SiO2data}. The process of scanning and digitizing the experimental plots was performed with data digitalization software \footnote{The data digitalization software is Plot Digitizer \url{https://plotdigitizer.com/}}.

The viscosity data inherently display noise present in the original experiments and scanning errors. In order to reduce the noise, we applied equidistance interpolation to the data and designed a low-pass Finite Impulse Response (FIR) filter \cite{proakis2001digital, parker2010digital, lyons2010digital} (see Appendix \ref{lowpassfilter}). We then fitted the filtered data with \myref{eq:2} for all temperatures above the liquidus temperature $T_{l}$ as well as the scaling temperature $T_{sc}$ discussed in the Introduction.

\begin{table*}[t]
	\centering
	\caption{Measurements of liquidus temperature $T_l$, crossover temperature $T_{sc}$ by us and $T_A$ from previous published work\cite{Tcoop1}, $\ln\eta_0$ and $\ln(nh)$ (both are measured in $Pa \cdot s$)\cite{Tcoop1}. For metallic liquids, their $\eta_0$ is nearly equal to their $nh$ value (see \myref{eta0nh}). By contrast, the liquids $\ce{H2O}$ and $\ce{B2O3}$ have $\eta_0$ values that are much smaller than their respective $nh$ values. The error bar of $T_{sc}$ is $1K$ while the error bar of $\ln\eta_0[Pa\cdot s]$ is $0.001$. The method of determining $T_{sc}$ (and thus also the corresponding $\ln\eta_0$) is provided in Fig. \ref{fig:zrptraw0}. Where the data was taken from sources other than our own measurements and/or \cite{Tcoop1}, these have been cited below. The data of \ce{H2O} have been calibrated by combining \cite{water,newwater}. The column marked  ``TB bound'' is a lower bound on $\ln\eta$ \cite{newwater}, which we will return to in Section \ref{sec:comparelnnh}. When the parameters were unknown they were left blank.}
	\vspace{1mm}
	\renewcommand\arraystretch{1.2}
	\tabcolsep0.14in
	\begin{tabular}{llllllll}
		\hline
		Composition&				$T_l$&		$T_{sc}$&	$T_{A}$&	$\ln\eta_0$&		$\ln(nh)$&		TB bound&		Density at $T_l$\cr
		&					$[K]$&		$[K]$&		$[K]$&		$[Pa\cdot s]$&		$[Pa\cdot s]$&		$[Pa\cdot s]$&		$[g/cm^3]$\cr
		\hline
		\ce{Cu43Zr45Al12}&			1209&		1371&		&		-10.829 \cr
		\ce{Cu46Zr54}&			1198&		1212&		&		-10.601 \cr
		\ce{Cu47Zr45Al8}&			1190&		1345&		&		-10.868 \cr
		\ce{Cu47Zr47Al6}&			1172&		1307&		&		-10.756& 		&			&			6.83\cr
		\ce{Cu49Zr45Al6}&			1177&		1324&		&		-10.945\cr
		\ce{Cu50Zr40Ti10}&			1168&		1276&		&		-10.877& 		&			&			6.90\cr
		\ce{Cu50Zr42_{.5}Ti7_{.5}}&		1152&		1237&		&		-10.903& 		&			&			6.92\cr
		\ce{Cu50Zr45Al5}&			1173&		1329&		1308&		-10.879& 		-10.2258&		-6.3847&		6.91\cr
		\ce{Cu50Zr50}&			1226&		1273&		1284&		-10.831& 		-10.2419&		-6.3796\cr
		\ce{Cu53Zr45Al2}&			1199&		1290&		&		-10.915\cr
		\ce{Cu55Zr45}&			1193&		1298&		&		-11.003\cr
		\ce{Cu60Zr20Ti20}&			1127&		1302&		1301&		-11.174& 		-10.0991&		-6.3168&		6.92\cr
		\ce{Cu60Zr40}&			1168&		1275&		&		-10.893 \cr
		\ce{Cu64Zr36}&			1230&		1320&		&		-11.139 \cr
		LM601&				1157&		1318&		&		-10.588 \cr
		\ce{Ni59_{.5}Nb40_{.5}}&		1448&		1637&		&		-10.479 \cr
		\ce{Ti38_{.5}Zr38_{.5}Ni21}&		&		1277&		&		-10.8\cr
		\ce{Ti40Zr10Cu30Pd20}&		1189&		1297&		1299&		-10.901& 		-10.1521&		&			6.82\cr
		\ce{Ti40Zr10Cu36Pd14}&	 	1185&		1274&		1278&		-10.952& 		-10.1360&		-6.3689&		6.69\cr
		Vit105&					1093&		1369&		&		-10.618& 		&			&			6.46\cr
		Vit106&					1123&		1362&		1373&		-10.505& 		-10.3156&		&			6.44\cr
		Vit106a& 				1125&		1357&		1360&		-10.646& 		-10.3156&		&			6.51\cr
		\ce{Zr57Ni43}&				1450&		1342&		&		-10.414\cr
		\ce{Zr59Ti3Ni8Cu20Al10}&		1145&		1313.5&	&		-10.691& 		-10.3547&		&			6.41\cr
		\ce{Zr60Ni25Al15}&			1248&		1395&		1421&		-10.516& 		-10.3662&		&			6.23\cr
		\ce{Zr62Cu20Ni8Al10}&			1152&		1321&		1325&		-10.531& 		-10.3639&		-6.5063\cr
		\ce{Zr64Ni25Al11}&			1212&		1350&		&		-10.394\cr
		\ce{Zr64Ni36}&				1283&		1256&		1223&		-10.27& 		-10.3271&		-6.4512\cr
		\ce{Zr65Al7_{.5}Cu17_{.5}Ni10}&	1170&		1267&		&		-10.274& 		&			&			6.5\cr
		\ce{Zr74Rh26}&			1350&		1357&		&		-10.144\cr
		\ce{Zr75_{.5}Pd24_{.5}}&		1303&		1289&		&		-10.284\cr
		\ce{Zr76Ni24}&				1233&		1157&		1161&		-10.057& 		-10.4123&		-6.5125\cr
		\ce{Zr80Pt20}\cite{Tcoop1, Zr80Pt20Density}&1450&	1482&		1549&		-10.004&		-10.3939&		-6.3468&		8.38\cr
		\ce{H2O} \cite{water,newwater}&  	273.15&	 297&		&	-13.4 &		-10.7181&		-9.6671&		0.997\cr
		\ce{B2O3} \cite{B2O3}&		723&		1187&		&		-4.289&			-11.1740&		-9.8872&		2.460\cr
		OTP \cite{OTPdata}&			330&		411&		&		-13.05\cr	
		T$\alpha$ NB \cite{OTPdata}&	435&		652&		&		-13.5\cr		
		\ce{SiO2}\cite{SiO2data}&		1986&		3455&		&		-3.4&			-11.136&		-9.7028&		2.65\cr	
		\hline
	\end{tabular}
	\label{tab:11}
\end{table*}

\section{Extracting the effective, temperature dependent, activation energies from numerical derivatives}
\label{sec:activation_barriers}

\begin{figure*}[t]
	\centering
	\includegraphics[scale=1]{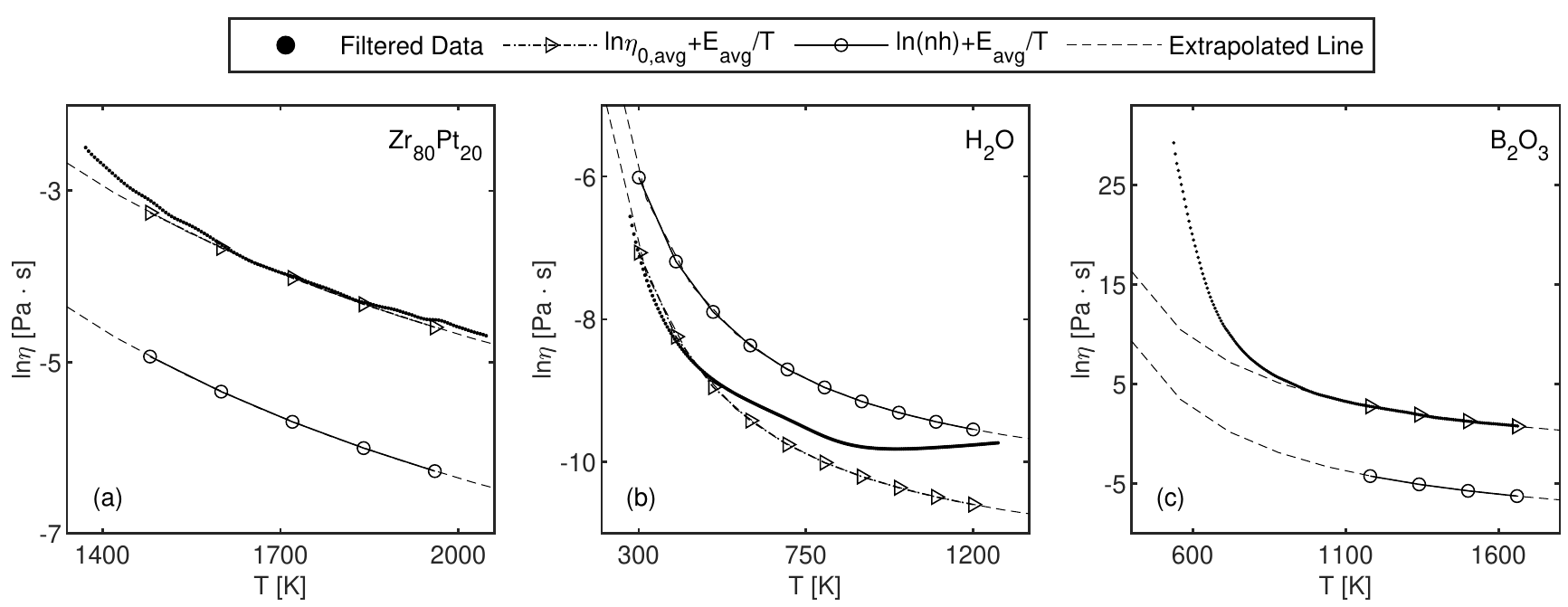}
	\caption{(Color online.) 
		Comparison between the measured viscosity of $\ce{Zr80Pt20}$, $\ce{H2O}$, and $\ce{B2O3}$ with the Arrhenius form of \myref{eq:2} (see Section \ref{sec:comparelnnh} for further comparisons). The ordinate represents the natural logarithm of the numerical value of the viscosity $\eta$ when the latter is measured in units of $Pa \cdot s$. 
		In all panels, the black scatters refer to the filtered experimental $\ln \eta(T)$ data that are contrasted with the Arrhenius fits of \myref{eq:2}. 
		 The line delineated by triangular markers is the Arrhenius fit of \myref{eq:2} obtained with an optimal uniform ``average'' (see Appendix \ref{s-c} for details) activation energy $E_{avg}$ with a viscosity prefactor $\ln \eta_{0,avg}$ values that fit the data well. 
		The curve with circular markers represents the fit of \myref{eq:2} by setting $\ln \eta_0=\ln(nh)$ where $n$ is the number density and $h$ is Planck's constant. (Similar to the logarithm of the viscosity, $\ln (nh)$ denotes the natural logarithm of the numerical value of $(nh)$ when $(nh)$ is measured in units of $Pa \cdot s$.) The thin dashed curves are extrapolations. 
		(a) The filtered experimental data of $\ce{Zr80Pt20}$ are compared with Arrhenius fits with $E_{avg}=8080K$, $\ln\eta_{0,avg}=-8.666$ and $\ln\eta_{0}=\ln(nh)=-10.394$ for temperatures above $T_l=1450K$. 
		(b) A comparison between the measured viscosity of $\ce{H2O}$ with the Arrhenius fit of \myref{eq:2}. An optimal fit for $T>T_{l} =273.15K$ is obtained by setting $E_{avg}=1411K$, $\ln\eta_{0,avg}=-11.770$ and $\ln(nh)=-10.718$. 
		(c) A comparison of $\ce{B2O3}$ with $E_{avg}=8189K$, $\ln\eta_{0,avg}=-4.177$ and $\ln(nh)=-11.174$ for $T>T_{l}=723K$. }
		\label{fig:lnnhlneatfit}
\end{figure*}

\begin{figure}[t]
	\centering
	\includegraphics[scale=1]{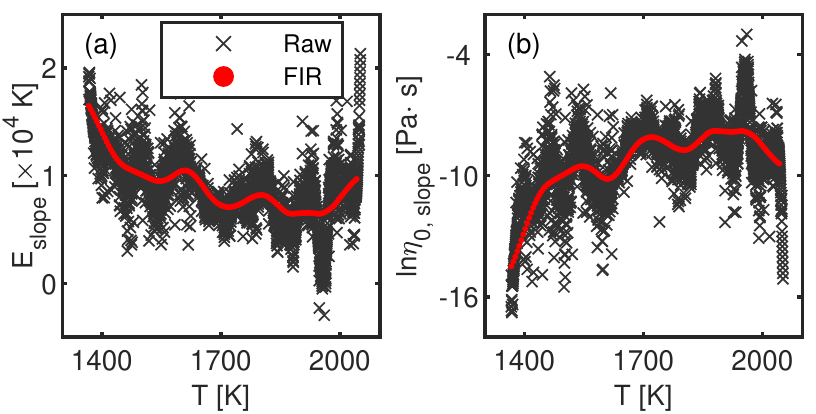}
	\caption{(Color online.)
	(a) The effective activation energy $E_{slope}$ of \myref{eq:60} for the metallic liquid $\rm Zr_{80}Pt_{20}$. The black curve is the raw finite difference gradients (\myref{eq:60} as evaluated over temperature windows of width $(T_{i+1} - T_{i})$ between consecutive points). The red curve displays these finite difference gradients as smoothened via a finite impulse filter (FIR)  (Appendix \ref{lowpassfilter}).  
	(b) The effective $\ln\eta_{0,slope}$ of the metallic liquid $\rm Zr_{80}Pt_{20}$ as evaluated over different temperature windows. Similar to (a), the black curve provides the raw finite difference gradients of \myref{eq:201} represents $\ln\eta_{0,slope}$ while the red curve corresponds to a finite difference gradient generated by an FIR with a larger temperature window.
	}
	\label{fig:zrptfir}
\end{figure}

Our filtered data are quite detailed with a minimal temperature interval $\Delta T=0.25K$ between subsequent data points (labelled an index $i$). As briefly alluded to in Section \ref{linear_invariance:sec}, this allows for numerical approximations of \myref{eq:4} via finite temperature differences, 
\begin{equation}
E_{slope}=\frac{\ln\eta_{i+1}-\ln\eta_i}{1/T_{i+1}-1/T_i}.
\label{eq:60}
\end{equation}
The finite difference approximation of \myref{eq:60} to the derivative of \myref{eq:4} is acutely sensitive to local variations of $\ln\eta$ as a function of the temperature. This drawback compounds the errors already present in the raw data. In order to see an intelligible trend, we apply a low-pass FIR filter (see Appendix \ref{lowpassfilter}) to calculate $E_{slope}$. As an illustrative example, we consider several aspects  of $\ce{Zr80Pt20}$ in Figures \ref{fig:lnnhlneatfit} and \ref{fig:zrptfir}. The black curve in Fig. \ref{fig:zrptfir}(a) represents the activation energy $E$ as obtained from \myref{eq:60} whereas the red curve is the resulting plot after applying the filter. We used this method to obtain $E$ as a function of $T$ above $T_l$ as displayed in Fig. \ref{fig: eeta0part}. (The analogous results for $T>T_{sc}$ appear in the Appendix (Fig. \ref{fig:e_eta0_s_metallic} and Fig. \ref{fig:e_eta0_s}.)

\section{Determining an effective $\eta_{0}$ in different temperature windows from numerical derivatives}
\label{sec:eta0}

Assuming a constant activation barrier, an effective temperature dependent $\ln\eta_0$ may, as we discussed earlier, be computed via \myref{eq:200}. Similar to \myref{eq:60}, the derivative in \myref{eq:200} may be approximated by a finite difference gradient,
\begin{equation}
\ln\eta_{0,slope}=\frac{(T\ln\eta)_{i+1}-(T\ln\eta)_i}{T_{i+1}-T_i}.
\label{eq:201}
\end{equation}

The prefactor results of Fig. \ref{fig: eeta0part} for temperatures above the liquidus temperature $T_l$ (and those of Fig. \ref{fig:e_eta0_s} of the Appendix for temperatures larger than a crossover temperature $T_{sc}$) illustrate, unambiguously, that the Arrhenius form does not hold. If $E$ is kept fixed then the prefactor $\eta_{0}$ of \myref{eq:1} cannot be a temperature independent constant; in most liquids, the value of $\eta_{0}$ necessary to fit the data changes by several orders of magnitude.

In Fig. \ref{fig:lnnhlneatfit}, we contrast the optimal values of the activation energy $E_{avg}$ with $E_{slope}$ (\myref{eq:60}) and the viscosity prefactor $\eta_{0,avg}$ with $\eta_{0,slope}$ of \myref{eq:201}. Here,  $E_{avg}$ and $\eta_{0,avg}$ refer to the constant (temperature independent) values of the activation energy and viscosity prefactor in the Arrhenius expression for the viscosity that fit the data best. Details of the optimization procedure are given in Appendix \ref{s-c}. As we noted above, according to  theoretical (Eyring-type) predictions \cite{eyring1,qmrelaxtime1,Tcoop1,kelton2016,otp1967}, the viscosity prefactor 
\begin{eqnarray}
\label{eta0nh}
\eta^{\sf theory}_{0} = nh. 
\end{eqnarray}
As we alluded to earlier, $n$ denotes the number particle density and $h$ is Planck's constant. In Table \ref{tab:11}, we compare the found fitted values with the prediction of \myref{eta0nh}. While the discrepancy between the empirical value of $\eta_{0}$ and $\eta^{\sf theory}_{0}$ is relatively small for {\it metallic liquids} \cite{qmrelaxtime1,Tcoop1} (see also Fig. \ref{fig:lnnhlneatfit}(a)), it can become far more marked for non-metallic fluids fitted over a large temperature range (\ref{fig:lnnhlneatfit}(b), (c)). The explicit functional form for the viscosity associated with the scaling temperature $T_{sc}$ will be elaborated on in \myref{Fcoll}.

\begin{figure}
	\centering
	\includegraphics[scale=1]{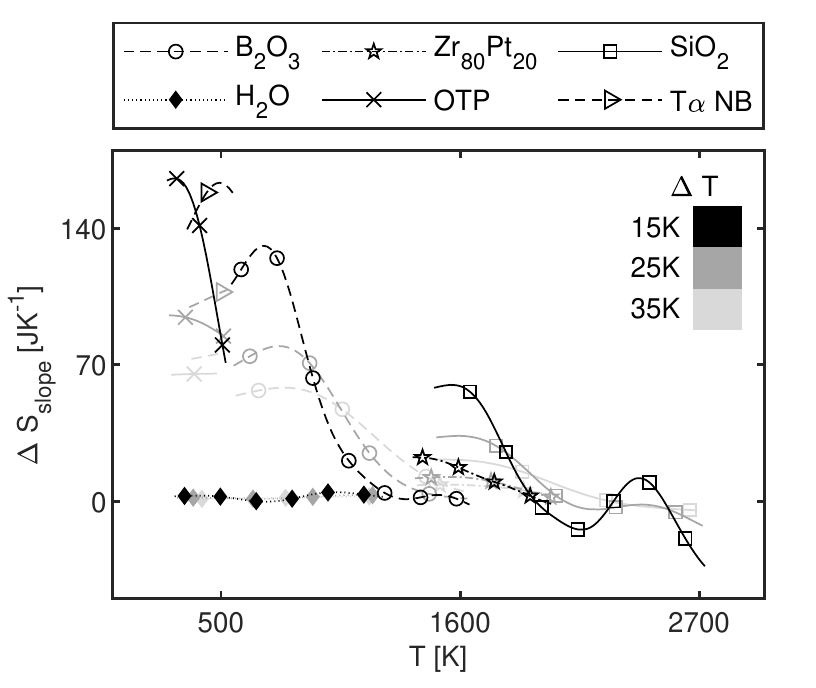}
	\caption{(Color online.) The effective entropy $\Delta S_{slope}$ defined by \myref{Sslope} for selected liquids above their liquidus temperature $T_{l}$. The finite difference gradients of \myref{Sslope} are evaluated over consecutive temperature windows ($i$) of width ($T_{i+1}- T_{i}$). Excusing fluctuations in the numerical data, the effective activation barrier generally decreases with increasing $T$. Thus, when computing numerical differences over a larger temperature interval, the average effective entropy as computed by \myref{Sslope} is positive. As the color bar indicates, the black curves are all defined by \myref{Sslope} at temperature interval $\Delta T=15K$, and the gray and light gray curves are similarly computed for, respectively, $\Delta T=25K, 35K$.}
	\label{fig: zrptentropy}
\end{figure}

\section{Effective Entropy}
\label{effEnt}

When employing the Eyring form of \myref{eyring:eq}, if the Gibbs free energy barrier $\Delta G$ is almost constant then Arrhenius dynamics will appear. Our results establish, however, that the effective activation barrier ($E_{slope}$) required to conform the experimental data is clearly temperature dependent. Thus, if we attempt to describe the data with the Eyring equation that the effective Gibbs free energy barrier $\Delta G$ varies with temperature. Such a variation implies that the {\it {effective entropy}} 
\begin{eqnarray}
\label{deltaS}
\Delta S \equiv - \Big(\frac{\partial \Delta G}{\partial T} \Big)
\end{eqnarray}
does not vanish.
To ascertain the scale of this effective entropy, we may replace $\Delta G$ by $E_{slope}$ and employ our finite temperature difference approximations (that we earlier invoked to determine $E_{slope}$) to rewrite \myref{deltaS} 
\begin{eqnarray}
\label{Sslope}
\Delta S_{slope} = -\frac{E_{slope,i+1}-E_{slope,i}}{T_{i+1}-T_i}.
\end{eqnarray}
In Fig. \ref{fig: zrptentropy}, we display both the filtered and unfiltered results of $\Delta S_{slope}$ resulting from such finite temperature differences (with, in \myref{Sslope}, the energies $E_{slope}$ explicitly measured in Joules (i.e., not, as in much of this work, rescaled by the $k_{B}$ and represented as a temperature scale)) when examining \ce{Zr80Pt20}. Employing a larger temperature interval $\Delta T$ avoids noise in the data and consistently yields positive effective entropy change $\Delta S$. This effective entropy change underscores the deviation from Arrhenius dynamics with a constant effective energy barrier. The positive sign of $\Delta S$ highlights, once again, the monotonic variation of the effective activation barrier with temperature in the liquid phase. Performing a linear fit illustrates that, on average, the ascertained effective entropy $\Delta S_{slope}$ monotonically rises with increasing temperature. The latter further implies
an average positive ``effective heat capacity'' $C_{eff} \equiv T \frac{\partial \Delta S}{\partial T}$.

\section{Tests of a more general universal viscosity collapse}
\label{collapse:sec}

\begin{figure*}[t]
	\centering
	\includegraphics[scale=1]{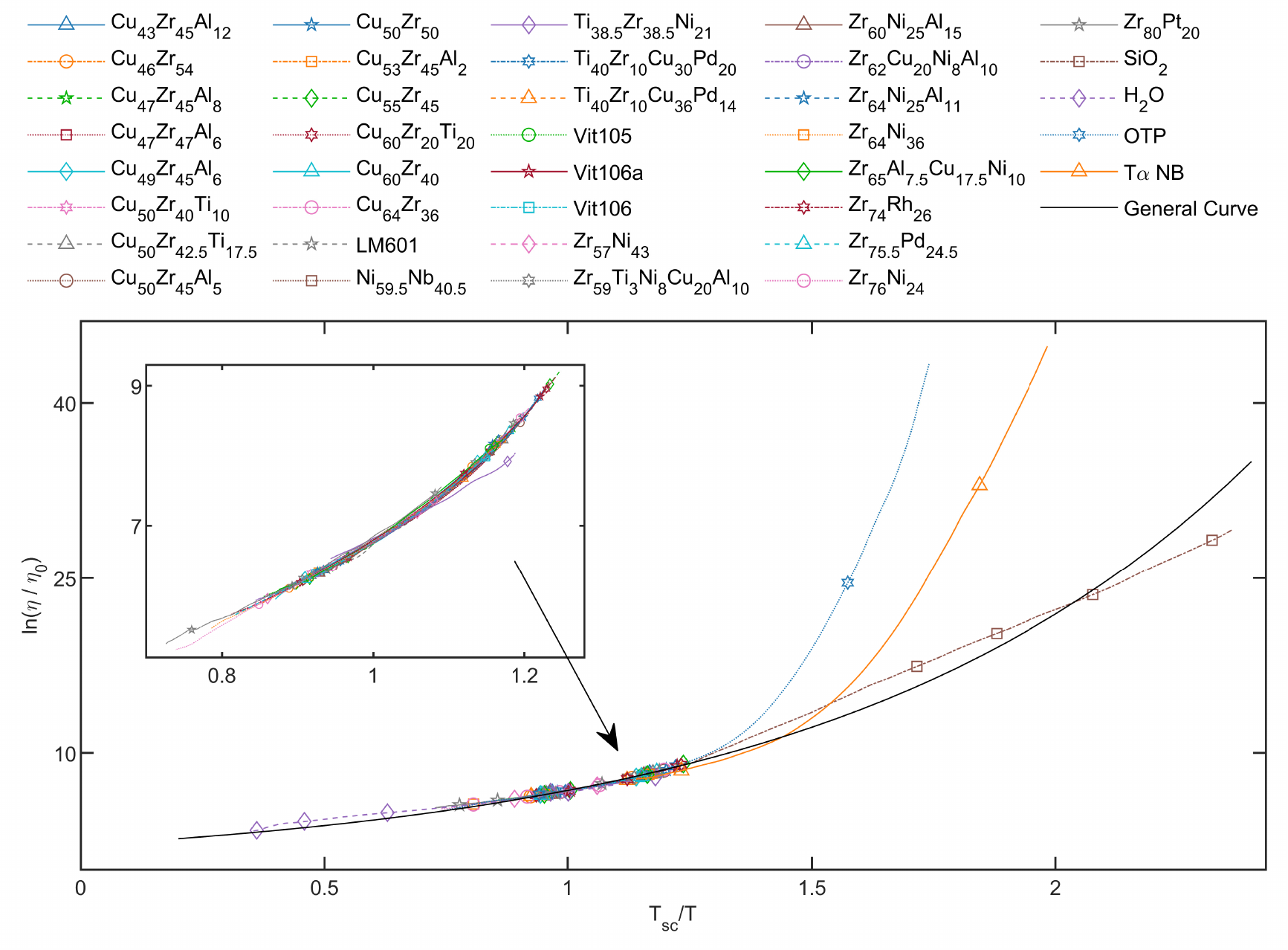}
	\caption{(Color online.) A test of a possible universal dimensionless collapse of the viscosity of liquids of different types (OTP, $\ce{H2O}$ and numerous metallic liquids). For $\ce{B2O3}$,  T$\alpha$NB, $\ce{SiO2}$ and OTP, the scaled curves collapse for $T_{sc}/T<1.25$. The values of $T_{sc}$ and $\eta_0$ associated with the displayed viscosity collapse are provided in Table \ref{tab:11}. The black dashed curve represents the collapse curve $\ln(\eta / \eta_0) = 2.111 e^{(1.19(T_{sc}/T))}$.}
	\label{fig:determineta}
\end{figure*}

\begin{figure}[h]
	\centering
	\includegraphics[scale=1]{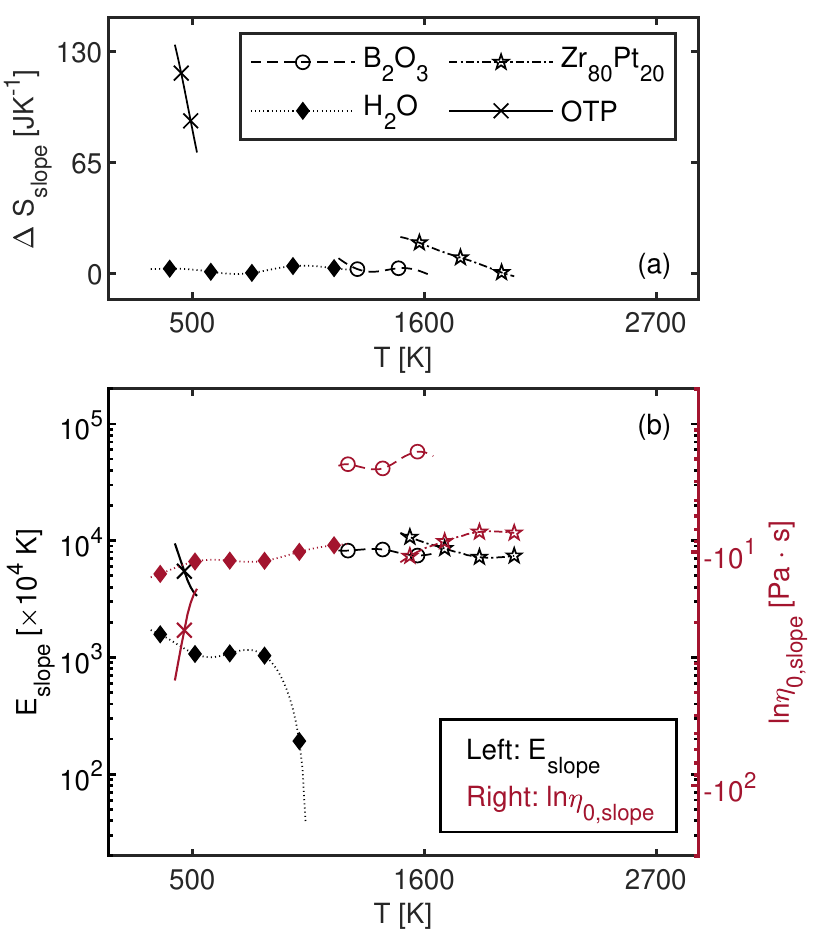}
	\caption{(Color online.) 
	(a) The effective entropy $\Delta S_{slope}$ of \myref{Sslope} and
	(b) the activation energy $E$ and the values of $\eta_{0}$ as a function of temperature above $T_{sc}$ for selected liquids except \ce{SiO2} and $T\alpha NB$ (see text) above their $T_{sc}$. For these two liquids, the values of $T_{sc}$ are higher than the examined temperature range and are thus not shown in the figure. }
	\label{fig:e_eta0_s_selected}
\end{figure}

Having established that viscosities may, generally, be far more complex than simplest activated functions of the temperature, we now ask whether more general scaling forms may better fit the data. That is, we will now inquire whether hallmarks of the commonly assumed universal (activated) dynamics may be seen by broader tests. To achieve this goal, we critically tested if $\eta(T)/\eta_{0}$ might be another (not necessarily the simple exponential appearing in the Arrhenius equation) universal function $F$ of a dimensionless temperature $T_{sc}/T$ with both $\eta_{0}$ and the scaling temperature $T_{sc}$ being specific constants for each fluid. If such a universal function exists then plotting, for $N$ disparate liquids, the dimensionless viscosity $\eta/\eta_{0}$ as a function of the dimensionless temperature $T_{sc}/T$ (with $T_{sc}$ replacing the activation barrier $E$ of the Arrhenius form) will lead to curve whose form is given by the aforementioned universal function $F$. As is well known, when present, a data collapse onto a universal curve underscores an underlying simplicity. The celebrated Guggenheim fit \cite{GuggenheimEA} first illustrated that the scaled dimensionless densities of various liquids in the vicinity of their critical points are a universal function of scaled dimensionless reduced temperatures. Guggenheim reported on this data collapse onto a universal curve long before the current advent of critical phenomena \cite{Stanley}. Inspired by these well known results, we assess to what extent a collapse might or might not occur for the viscosities of various fluids with such liquid dependent adjustable temperature ($T_{sc}$) and viscosity ($\eta_{0}$) scales, 
\begin{eqnarray}
\label{Fcoll}
\frac{\eta}{\eta_{0}} = F\Big(\frac{T_{sc}}{T} \Big).
\end{eqnarray} 
Here the function $F(z)$ {\it will not} be constrained to the exponential function ($e^{cz}$ with $c$ a constant) defining the Arrhenius form of \myref{eq:1}. Operationally, we adjust the constants $T_{sc}$ and $\eta_{0}$ such that, the scaled curves of $\ln (\eta/\eta_{0})$ as a function of $T_{sc}/T$ of the different fluids enjoy a large overlap. Earlier works  \cite{Tcoop1} examined the prospect of such adjustable scales particularly with regard to a possible crossover of viscosities of supercooled liquids from Arrhenius to super-Arrhenius dynamics (see our own analysis for one such glass former ($\rm Zr_{80}Pt_{20}$) in Fig. \ref{fig:zrptraw0} where the crossover temperature is marked as $T_A$).
We emphasize that it is because the Arrhenius form does not work well (as we illustrated in the previous sections) hat we test to see if the functional form of \myref{Fcoll} fits the viscosity data better. Since it includes the Arrhenius form as a special case), \myref{Fcoll} will always allow for a broader collapse of the data than when $F$ is constrained to an Arrhenius form. If one the resultant viscosity collapse does not extend over a significant range of scaled temperatures $(T/T_{sc})$ then the deviation from any attempted scaling of such a form will be even stronger (i.e., no collapse appears even if the function $F$ in \myref{Fcoll} is not restricted to be the exponential function associated with the Arrhenius fit). 
We determined the values of $T{sc}$ and $\eta_{0}$ by maximizing the overlap between the $\ln(\eta/\eta_{0})$ vs. $T_{sc}/T$ curves of the different fluids. Towards this end, we calculated, for any pair of fluids, the discretized integral of the squared difference (a sum of squared errors (SSE)) between scaled viscosity curves of the two fluids. We then found the values of scaling parameters $T_{sc}$ and $\ln\eta_0$ that minimized the resulting latter SSE when the latter was summed over all $N(N-1)/2$ pairs of fluids.

For any two different liquids, there is a specific SSE value. For 38 different liquids, there are 703 pairs of liquids and 703 SSE values. Taking the sum of all 703 SSE values, we obtain an overall SSE, which will vary with the change of each liquid's $T_{sc}$ and $\eta_0$. By adjusting $T_{sc}$ and $\eta_0$ for each liquid, we are able to minimize the overall SSE and to optimize our collapsed curve. The values of $T_{sc}$ and $\eta_0$ after the adjustment are therefore our optimum $T_{sc}$ and $\eta_0$ values. 

In Fig. \ref{fig:determineta}, we display our test results for a possible general viscosity collapse of high temperatures liquids with an unconstrained function $F$ in \myref{Fcoll} that is not, necessarily, of an Arrhenius form. As seen therein, the scaled viscosities of several liquids (e.g., $\ce{Cu50Zr40Ti10}$, OTP, $\ce{H2O}$, etc.) track each other over many decades. The silicate $\ce{SiO2}$, a quintessential ``strong'' glass former with relatively small deviations from Arrhenius dynamics upon supercooling to low temperatures, also displays somewhat minute differences from activated dynamics at high temperatures relative to the other liquids that we examined. The viscosity of some fluids may be collapsed in this way onto one another over many decades of the viscosity while for other fluids an attempted collapse yields more fleeting results- a viscosity collapse does not appear for all liquid types in unison. The values of $T_{sc}$ that we found for various fluids are often close to the values of $T_A$ (see Table \ref{tab:11}) associated with deviations from approximate high temperature Arrhenius dynamics.

\section{A Lower scale bound on the viscosity}
\label{sec:comparelnnh}

\begin{figure}[t]
	\centering
	\includegraphics[scale=1]{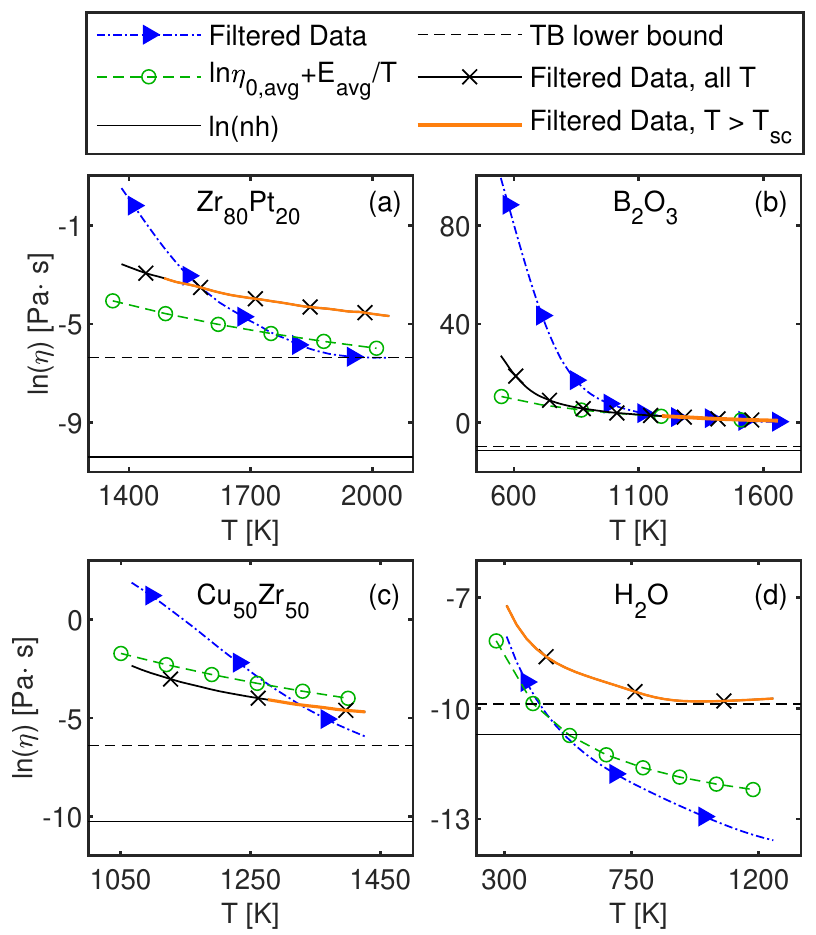}
	\caption{(Color online.) Comparison of different lower bounds of the viscosity for (a) \ce{Zr80Pt20}, (b) \ce{B2O3}, (c) \ce{Cu50Zr50}, and (d) \ce{H2O} above their liquidus temperature $T_l$. The black only portion of the continuous curve is comprised of filtered data points having temperature  $T>T_l$ while the overlaid orange portion of that curve provides the viscosity in the temperature range $T>T_{sc}$. In (d), the data range is such that these black and orange data points completely overlap. The bottom solid horizontal line marks the value of $\ln(nh)$ whose scale is a proposed lower bound of viscosity  \cite{qmrelaxtime1,Tcoop1,hbound} and the dashed line is a more recent bound suggested by \cite{newwater} (Eq. (\ref{con:ktlowerbd})). The blue curve represents the Arrhenius form with temperature dependent $\eta_{0,slope}$ and $E_{slope}$, while the green dashed curve refers to the Arrhenius form with constant $\eta_{0,avg}$ and $E_{avg}$. In these and all other fluids that we investigated, the raw viscosity data was consistent with the two viscosity bounds of Eqs. (\ref{lowernh}, ref{con:ktlowerbd}). 
	}
	\label{fig:lowbd}
\end{figure}

As we briefly reviewed in the Introduction, the viscosity of various compounds is typically minimal at a crossover between their gaseous and fluid phases. Several investigations \cite{qmrelaxtime1,Tcoop1,newwater,hbound} suggested a lower bound on this viscosity minimum. In this Section, we discuss two interrelated bounds and illustrate that they are satisfied for the liquids that we examined. The first bound is that of Eqs. (\ref{lowernh}, \ref{eta0nh}) \cite{qmrelaxtime1,Tcoop1,hbound}. A related second bound, proposed by \cite{newwater} (TB), can be expressed as
\begin{equation}
\eta \ge \frac{nh}{8\pi^2} \sqrt{\frac{m}{m_e}}.
\label{con:ktlowerbd}
\end{equation}
Here, $m_e$ is the electron mass, and $m$ is the mass of the molecules forming the liquid. With $M=(m/(1836 m_{e}))$ denoting the molecular mass of the fluid, the bound of \myref{lowernh} is lower by a factor of $\sim 0.543 \sqrt{M}$ relative to the TB bound of \myref{con:ktlowerbd} (dashed pink line). In Fig. \ref{fig:lowbd}, we tested these bounds against available experimental data. For \ce{H2O}, the viscosity minimum at $800$ K saturates the TB bound. In Fig. \ref{fig:lowbd}, we further include, for comparison, two extended Arrehnius type forms (one with the temperature dependent $E_{slope}$ and $\eta_{0,slope}$ and the other with temperature independent $E_{avg}$ and $\eta_{0,avg}$) shown in blue and green. Both of these Arrhenius type functions deviate substantially from the measured viscosity curve. These deviations underscore the invalidity of the Arrhenius form for describing the viscosity of these systems. Over the temperature range shown for water, a strong deviation from Arrhenius is mandated since the viscosity rises with increasing temperature in sufficiently high temperature gases. The other three systems displayed in Fig. \ref{fig:lowbd}  (Zr$_{80}$Pt$_{20}$, B$_2$O$_3$, and Cu$_{50}$Zr$_{50}$) are all far below their respective boiling temperatures.

\section{Conclusions}
\label{sec:conclusions}
We tested the validity of Arrhenius form for describing the dynamics of general liquids at temperatures above those of melting (and other possible crossovers) by carefully analyzing viscosity data and contrasting it with \myref{eq:1}. We applied an equidistant interpolation of the data and partitioned the temperature range into equal intervals. Subsequently, we applied a low-pass FIR filter to reduce the data noise. We computed the values of the putative uniform activation energy $E$ (and prefactor $\eta_0$) of \myref{eq:1} at $0.25K$ temperature intervals. 

$\bullet$ Our analysis indicates that the viscosity of the liquid (at all temperatures within that phase) is far more complex than a simple Arrhenius behavior with a single temperature independent activation energy $E$. Perusing Fig. \ref{fig: eeta0part} and ensuing analysis, one sees that the viscosity data may be qualitatively captured by the likes of \myref{eyring:eq} when, as a general trend the Gibbs free energy activation barrier $\Delta G$ typically increases as the temperature $T$ decreases. Equivalently, dispensing with local (in temperature) fluctuations, the associated {\it effective entropic contribution $\Delta S = - \frac{\partial \Delta G}{\partial T}$ is generally large and positive}. These trends are highlighted in Figure \ref{fig: zrptentropy}. 

$\bullet$ We tested whether activated (more general than Arrehnius) dynamics might still appear universally in high temperature liquids. Towards that end, we examined the extent to which it is possible to collapse dimensionless viscosity data of different fluids as a function of a scaled dimensionless temperature and found that several fluids (of very different composition) exhibit strikingly similar behaviors over many decades of viscosity while others are more divergent. We caution that our results and analysis concern only the viscosity. {\it We do not exclude possible Arrhenius behaviors of other transport coefficients.} 

$\bullet$ We found that the scale of the viscosity of metallic fluids is consistent with that provided by \myref{eta0nh} with $n$ being the particle number density and $h$ Planck's constant. More generally, we find that the lower bound scale of \myref{lowernh} holds empirically in both the metallic and non-metallic fluids that we examined.

In the Appendix, we further contrast the empirical viscosity data with several earlier fits in the literature (that were largely introduced for various glass formers). While the most prevalent fits assume a constant Arrhenius form at high temperatures, some do not. In particular, the MYEGA form  \cite{myega1, myegazq}, 
\begin{eqnarray}
\label{ref:MYEGA}
\ln\eta=\ln\eta_0+\frac{K'}{T}e^{C/T}
\end{eqnarray}
(with material dependent parameters $\eta_0, K',$ and $C$), 
and the DHTDSJ fit \cite{dhtdsj1},
\begin{eqnarray}
\label{ref:DHTDSJ}
\ln\eta=\ln\eta_0+\frac{W_0}{k_BT}e^{-T/T_W}
\end{eqnarray} 
(with its fluid dependent parameters $\eta_0,W_0,$ and $T_W$), 
may both be expressed as scaling exponentially in $E/T$ with effective energy barriers $E(T)$ that (unlike the Arrhenius form) increase as the temperature $T$ is decreased. The more rapid rise of the viscosity than predicted by activated dynamics dominated by a uniform energy barrier is a feature that we find in all liquids. In accord with these data trends, the MYEGA and DHTDSJ fits that allow (with adjustable additional parameters) for effective activation energies to become larger as the temperature drops fit the viscosity data better than the Arrhenius form that, as we demonstrated in the current work, exhibits sizable variations from the experimental data. There are various possible extensions of our extensive analysis of the viscosity of disparate fluids that formed the focus of the current study to disparate response functions- e.g., tests of the Arrhenius and Eyring forms for dielectric relaxation rates. Additionally, the relation between our findings regarding the temperature dependence of the effective activation behavior in equilibrated high temperature liquids and the far more dramatic ``super-Arrhenius'' viscosity of supercooled liquids \cite{Langer2014,supercool1996,supercool2,Angell1924,relaxAngell,qmrelaxtime1,kelton2016,correlation2016,relaxArr2012, berthier2011, BerthierBiroli2011} would be interesting to explore. 

We conclude with a more speculative remark. Following recent elegant analysis by Bagiolli and Zaccone \cite{BZ1,BZ2}, the density of states in the liquid is given by 
\begin{eqnarray}
\label{gliquid}
g(\omega) \sim \frac{\omega}{\omega^2 + \Gamma^2} e^{-\omega^{2}/\omega_{D}^{2}},
\end{eqnarray}
with $\omega_{D}$ an effective Debye frequency (such that the last factor introduces a soft cutoff) and $\Gamma$ a temperature dependent damping rate constant. As pointed out by \cite{BZ1}, this enables the computation of thermodynamic observables such as, e.g., the specific heat contribution from these instantaneous normal modes \cite{BZ1},
\begin{eqnarray}
\label{cvliq}
c_v = k_{B} \int_{0}^{\infty} d \omega ~  \frac {g(\omega)\frac{\hbar \omega}{2 k_{B} T}}{\sinh^2 \frac{\hbar \omega}{2 k_{B} T}}.
\end{eqnarray}
Given our findings in the current work of deviations from activated dynamics in general fluids, instead of assuming that $\Gamma$ obeys an Arrhenius type behavior \cite{BZ1}, we may attempt to, more generally, set $\Gamma$ equal to the reciprocal of the temperature dependent measured relaxation time i.e., $\Gamma = \tau^{-1}$, with $\tau$ determined by the Maxwell relation of \myref{eq:99}. This may fortify \cite{BZ1,BZ2} so as to afford a general link between dynamics ($\tau$) and thermodynamics ($c_v$) in fluids whose deviation from Arrhenius dynamics is marked over the pertinent temperature range. While such a link seems logical, caution must be paid to the contributions of the different frequencies to the dynamic and thermodynamic properties. The order of magnitude inconsistencies suggest that additional frequency contributions other than those in \myref{gliquid} may need to be included in $g(\omega)$. Towards this end, we comment on a discrepancy encountered when assuming the single damping rate (and Debye frequency) term of \myref{gliquid} to capture both the heat capacity and viscosity data at different temperatures. Indeed, the viscosity of liquids may be typically dominated by low frequency hydrodynamic modes whereas the heat capacity can be controlled by far stiffer short range Debye type elastic modes. For instance, for water, the latter hydrodynamic damping rate $\Gamma$ is $10^{-1}$ Hertz whereas the scale of the frequency with dominant contributions to the heat capacity is $10^{13}$ Hertz \cite{CalebS2022}. We leave the detailed analysis of the above suggested link to a future investigation.

\section{Acknowledgments}
We gratefully acknowledge support by NSF grant DMR 1411229 (ZN) which has since been terminated and NSF grant DMR 1904281 (KFK). We further wish to thank the Aspen Center for Physics (supported by NSF PHY-1607611) where some of this work was written.

\appendix

\section{Self-consistency checks}
\label{s-c}

\begin{figure*}[t]
	\centering
	\includegraphics[scale=1]{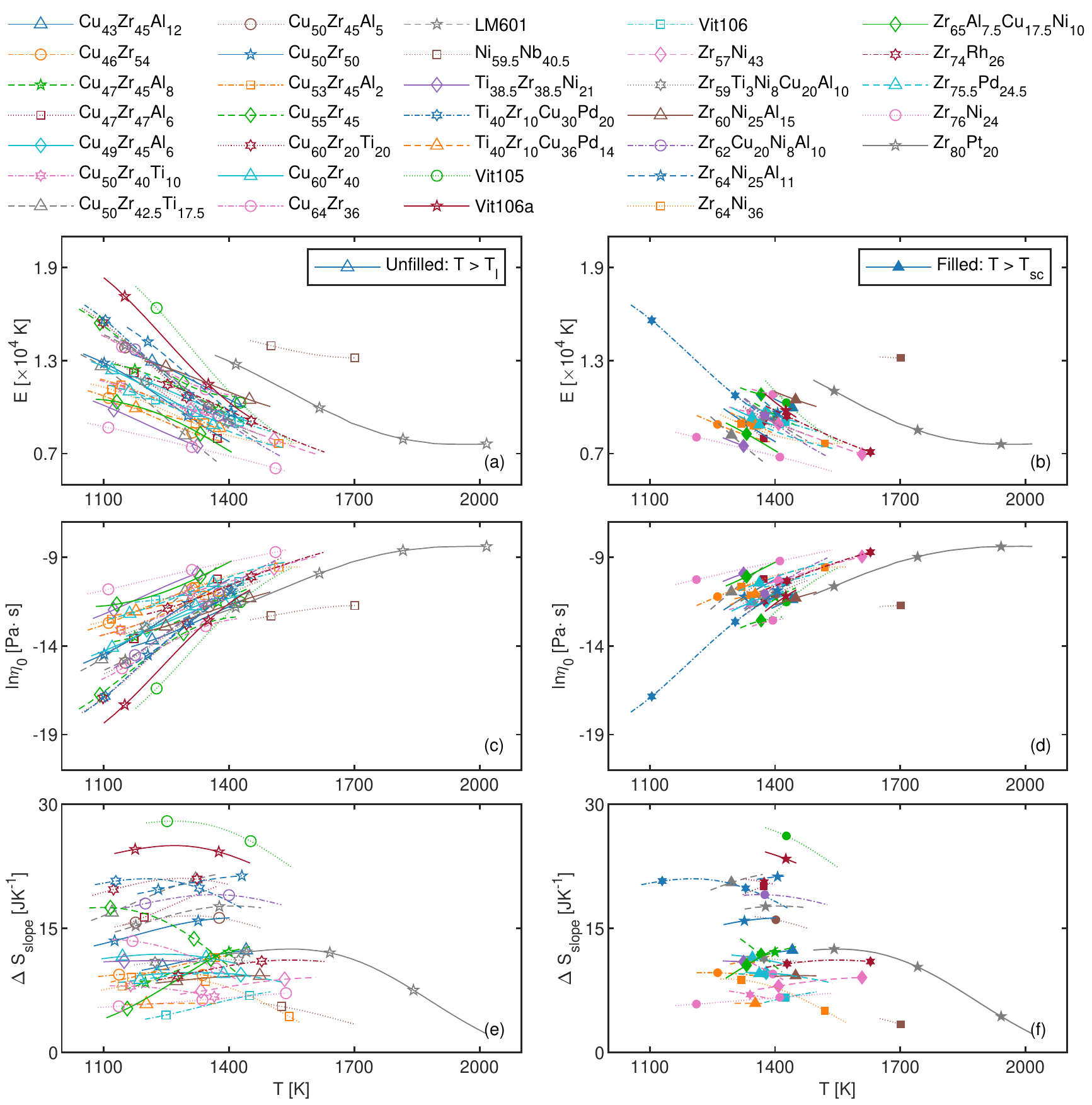}
	\caption{(Color online.) Activation energy $E$ (\myref{eq:4}), viscosity prefactor $\eta_{0}$ (\myref{eq:200}), and effective entropy $\Delta S_{slope}$ of \myref{Sslope} as a function of temperature of various fluids above their liquidus temperature $T_{l}$ (unfilled markers) and their scaling temperature $T_{sc}$ (filled markers).
	}
	\label{fig:e_eta0_s_metallic}
\end{figure*}

\begin{figure*}[t]
	\centering
	\includegraphics[scale=1]{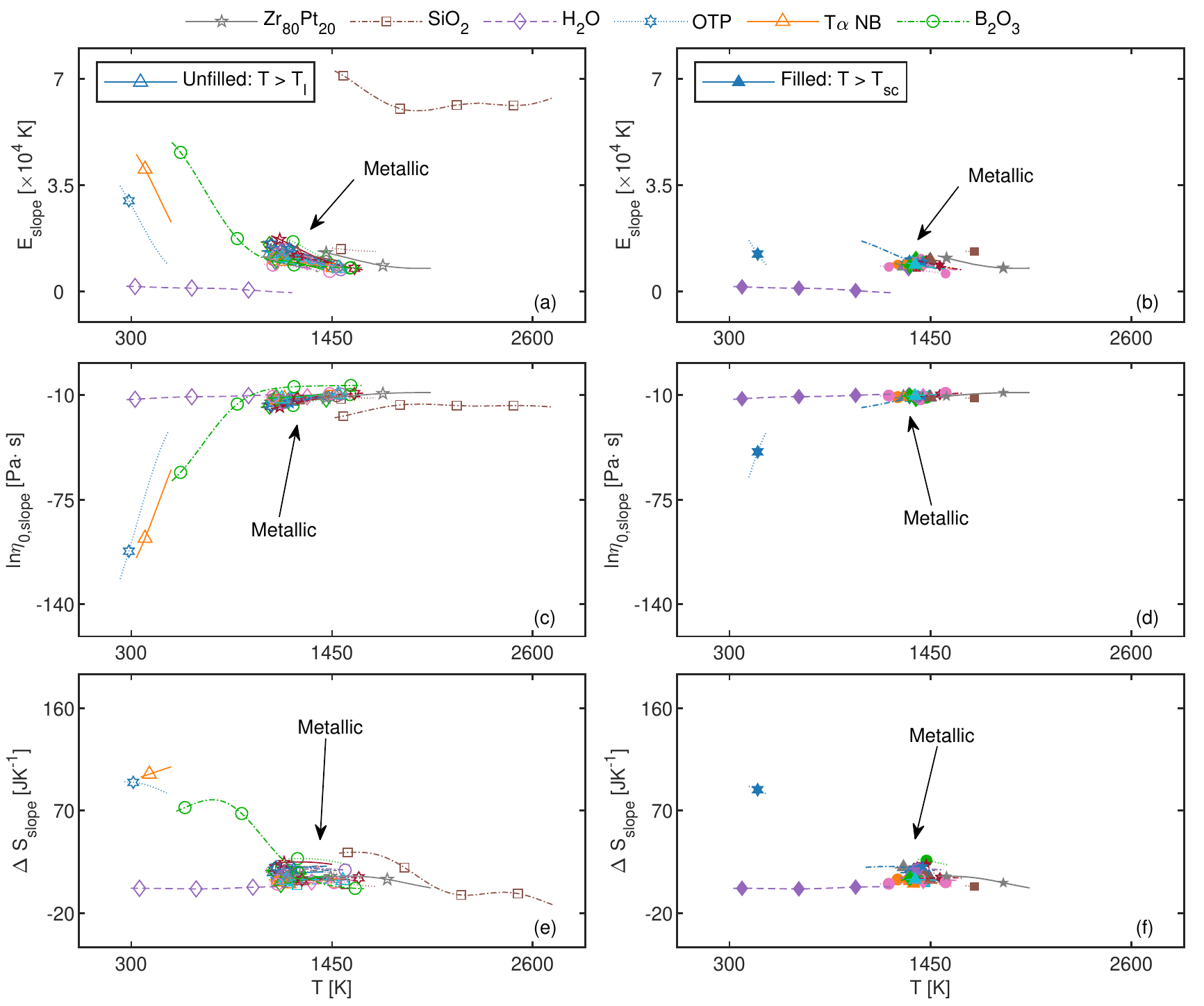}
	\caption{(Color online.) Similar to Figure \ref{fig:e_eta0_s_metallic}, we plot the activation energy $E$ (\myref{eq:4}), viscosity prefactor $\eta_{0}$ (\myref{eq:200}), and the effective entropy $\Delta S_{slope}$ of \myref{Sslope} as a function of temperature of several liquids above their liquidus temperature $T_{l}$ (unfilled markers) and their scaling temperature $T_{sc}$ (filled markers). Related results are shown in Fig. \ref{fig:e_eta0_s_selected} and Fig. \ref{fig:e_eta0_s_metallic}. Here, we provide data for additional fluids. For \ce{SiO2} and T$\alpha $NB (see text), the values of the scaling temperature $T_{sc}$ are higher than the examined temperature range and are therefore not shown in the figure. 
	}
	\label{fig:e_eta0_s}
\end{figure*}

It is illuminating to test the Arrhenius form by reinserting our obtained $E$ and $\ln\eta_0$ into \myref{eq:2}.  If the Arrhenius equation is valid then, up to reasonable scatter in the data, the activation energies $E(T)$ (and associated prefactors $\eta_{0}$) will assume constant ``average'' values $E_{avg}$ (and $\eta_{0,avg}$). To compute $E_{avg}$ (and $\eta_{0,avg}$), we take the equal weight uniform average of $E_{slope}$ (and $\eta_{0,slope}$) over the temperature range which is from $T$ above $T_{sc}$ to the maximum temperature of our liquids data range.

We may then substitute these average values $E_{avg}$ and $\ln\eta_{0,avg}$ into \myref{eq:2} and compare $\ln\eta=\ln\eta_{0,avg}+E_{avg}/T$ with the experimental data and contrast the so obtained $\ln\eta_{0,avg}$ from the experimental data with the theoretical prediction of \myref{eta0nh}. In Fig. \ref{fig:lnnhlneatfit}(a), an Arrhenius fit with $E_{avg}$ and $\ln\eta_{0,avg}$ (marked in red (color online)) is consistent with the raw data (the black curve in this Figure) only at temperatures close to $T_{sc}=1482K$. The Arrhenius curve and the actual raw experimental data substantially deviate from one another when extrapolating to higher (and lower) temperatures. We next tested how an agreement with the Arrhenius form might be ameliorated if we compute the average values $E_{avg}$ and $\ln\eta_{0,avg}$ over a narrower temperature range. Towards this end, we calculated the above $E_{avg}$ and $\ln\eta_{0,avg}$ by averaging  over the temperature interval between $T_{A\ast}$ to $T_{max}$ (where $T_{A\ast}>T_{sc}$). Here, $T_{max}$ denotes the highest temperature for which experimental data are available. Evaluating these averages, we found that the data and Arrhenius form with the above values matched in a limited range near $T_{A\ast}$. As we progressively shortened the temperature range over which the averages were taken (by fixing $T_{max}$ and raising $T_{A\ast}$), the Arrhenius (red) curve in Fig. \ref{fig:lnnhlneatfit}(a) continued to deviate from the experimental data at gradually higher temperature. 

We further explored the consistency of \myref{eta0nh} with the experimental data, where $E_{avg}=8225 K$ and $\ln (nh)=-10.394  Pa\cdot s$ for $\rm Zr_{80}Pt_{20}$. As Fig. \ref{fig:lnnhlneatfit}(a) underscores, an order of magnitude disparity may appear between \myref{eta0nh} and the experimental data - the scale of $\eta_{0}$ associated with the fitted measured viscosity of $\rm Zr_{80}Pt_{20}$ is, approximately, $e^{1.6} \sim 5$ times larger than the product $nh$. As seen in Fig. \ref{fig:lnnhlneatfit}(b), for $\ce{H2O}$ the corresponding ratio between $\eta_{0}$ and $(nh)$ is far larger, being approximately $e^6 \sim 400$. Due to measurement errors and the form of the raw experimental data, the temperature intervals may not, generally, be judiciously chosen so as to be of uniform width. Consequently, we cannot collate all of the calculated slopes into one figure to see how they vary with the temperature since each slope has different denominator values when employing Eqs. (\ref{eq:60}, \ref{eq:201}).

\section{Reliability of results}
\label{ReliabilityofResults}

\begin{table*}[t]
	\centering
	\normalsize
	\caption{Statistical residuals of fitting results. The first four columns are the residuals when fitting the raw and filterd data with Arrhenius form \myref{eq:200} with the whole data range, and the 4th and 5th coulumns are the $R^2$ results when fitting $T<T_A$ and $T>T_A$ with myref{eq:200}.	The uncertainty of each liquid comes from the liquids' density's liquids.The uncertainty of $E$ is determined by evaluating the maximum amplitude of the oscillated curve.}
	\vspace{1mm}
	\renewcommand\arraystretch{1.2}
	\tabcolsep0.1in
	\begin{tabular}{llllllll}
		\hline
		Composition&			SSE (raw)&		$R^2$(raw)&	    	SSE (filter)&	$R^2$(filter)&	$R^2$(filter)&		$R^2$(filter)&		Uncertainty\cr
		&				$[10^{-5}]$&		(All T)&			$[10^{-5}]$&	(All T)&		$(T<T_A)$&		$(T>T_A)$&		of $E$[$\times10^4 K$]\cr
		\hline
		\ce{Cu43Zr45Al12}&		1.576&			0.9502&		3.863& 		0.9998&	&			&			0.2\cr
		\ce{Cu46Zr54}&		8.125&			0.9489&		2.123& 		0.9982&	&			&			0.25\cr
		\ce{Cu47Zr45Al8}&		4.016&			0.9339&		1.908& 		0.9943&	&			&			0.3\cr
		\ce{Cu47Zr47Al6}&		5.112&			0.894&			2.214& 		0.9857&	&			&			0.2\cr
		\ce{Cu49Zr45Al6}&		1.237&			0.9735&		5.121& 		0.9987&	&			&			0.25\cr
		\ce{Cu50Zr40Ti10}&		4.522&			0.9032&		0.936& 		0.9972&	&			&			0.2\cr
		\ce{Cu50Zr42_{.5}Ti7_{.5}}&	1.518&			0.9874&		6.006& 		0.9947&	&			&			0.25\cr
		\ce{Cu50Zr45Al5}&		1.512&			0.9727&		3.214&		0.9936&	0.9912&		0.9918&		0.2\cr
		\ce{Cu50Zr50}&		3.231&			0.9281&		2.489&		0.9951&	0.9941&		0.9948&		0.3\cr
		\ce{Cu53Zr45Al2}&		3.667&			0.9465&		1.844& 		0.9972&	&			&			0.3\cr
		\ce{Cu55Zr45}&		3.941&			0.9016&		2.841&		0.9906&	&			&			0.2\cr
		\ce{Cu60Zr20Ti20}&		1.254&			0.9612&		0.609&		0.9955&	0.9836&		0.9964&		0.2\cr
		\ce{Cu60Zr40}&		5.126&			0.9189&		4.263& 		0.9966&	&			&			0.2\cr
		\ce{Cu64Zr36}&		3.12&			0.8317&		2.607& 		0.9834&	&			&			0.2\cr
		LM601&			1.039&			0.9665&		11.03&		0.9883&	&			&			0.2\cr
		\ce{Ni59_{.5}Nb40_{.5}}&	1.924&			0.9699&		9.659& 		0.9756&	&			&			0.2\cr
		\ce{Ti38_{.5}Zr38_{.5}Ni21}&	 4.199&			0.9839&		5.615& 		0.9969&	&			&			0.2\cr
		\ce{Ti40Zr10Cu30Pd20}&	2.653&			0.9496&		0.652&		0.9986&	&			&			0.25\cr
		\ce{Ti40Zr10Cu36Pd14}&	3.162&			0.9513&		0.7& 		0.9977&	0.9995&		0.9951&		0.3\cr
		Vit105&				4.566&			0.9732&		2.049&		0.9983&	&			&			0.2\cr
		Vit106&				0.001098&		0.9873&		0.00274&	0.9951&	0.9982&		0.9996&		0.2\cr
		Vit106a& 			2.585&			0.9546&		1.623&		0.9962&	0.9886&		0.9965&		0.2\cr
		\ce{Zr57Ni43}&			18.89&			0.976&			5.535&		0.998&		&			&			0.2\cr
		\ce{Zr59Ti3Ni8Cu20Al10}&	2.797&			0.9585&		1.694& 		0.9977&	&			&			0.2\cr
		\ce{Zr60Ni25Al15}&		1.201&			0.9847&		1.896& 		0.9969&	&			&			0.2\cr
		\ce{Zr62Cu20Ni8Al10}&		2.34&			0.9866&		 1.563&		0.9928&	&			&			0.2\cr
		\ce{Zr64Ni25Al11}&		6.04&			0.8718&		0.908& 		0.998&		&			&			0.3\cr
		\ce{Zr64Ni36}&			4.211&			0.9015&		1.788&		0.9972&	0.9966&		0.9985&		0.3\cr
		\ce{Zr65Al7_{.5}Cu17_{.5}Ni10}& 12.01&		0.9551&		15.42& 		0.9915&	&			&			0.2\cr
		\ce{Zr74Rh26}&		232.4&			0.9866&		564& 		0.9929&	&			&			0.15\cr
		\ce{Zr75_{.5}Pd24_{.5}}&	232.4&			0.9843&		998.2& 		0.991&		&			&			0.2\cr
		\ce{Zr76Ni24}&			19.13&			0.9919&		195.7& 		0.9983&	0.9999&		0.9994&		0.2\cr
		\ce{Zr80Pt20}&			2.001&			0.9913&		0.997&		0.9984&	0.997&			0.9935&		0.35\cr
		\ce{H2O}&			17.5&			0.9233&		2.45&		0.9932&	&			&			0.2\cr
		\ce{B2O3}&			4.017&			0.9542&		1.066&		0.9961&	&			&			0.1\cr
		OTP&				3.018&			0.9332&		1.656&		0.9947&	&			&			0.3\cr
		T$\alpha$ NB&			2.89&			0.9611&		1.788&		0.9936&	&			&			0.2\cr
		\ce{SiO2}&			1.526&			0.9991&		0.956&		0.9977&	&			&			0.1\cr
		\hline
	\end{tabular}
	\label{tab:22}
\end{table*}

We need to assess whether our obtained effective $E(T)$ and $\eta_0(T)$ are reliable. Addressing this question requires us to find the extent to which the FIR filter may impact the determined $E(T)$ and $\eta_0(T)$. The defining property of our (and any) FIR filter is the absence of feedback in its application generally endowing it with an intrinsic stability. 

In the main text, we reported on our tests of the validity of Arrhenius form by plotting both the activation energy $E(T)$ and the prefactor $\eta_0(T)$ as functions of temperature. Trends in $E(T)$ and $\eta_0(T)$ become clearer when these are smooth and monotonous. However, if we apply the finite difference equations of \myref{eq:60} or \myref{eq:201} directly to the raw viscosity data, then $E(T)$ or $\eta_0(T)$ will, generally, exhibit large fluctuations. These fluctuations can obscure trends in either the activation energy $E$ or the viscosity prefactor $\eta_0(T)$.

Ideally, reducing the width of the temperature interval in \myref{eq:60} and \myref{eq:201} may, in principle, allow for a more accurate determination of an effective activation energy $E(T)$ or the viscosity prefactor $\eta_0(T)$.
However, the use of smaller intervals will, inevitably, enhance any noise present in the data. Due to the measurement errors and the detailed form of the raw experimental data, generally, the temperature intervals may not chosen to identically be equal. To deal with the problem of varying width temperature intervals, we applied equidistant interpolation. To mitigate the inherent noise in the data, we designed and applied a low-pass FIR filter with equidistant interpolation. 

We applied the FIR filter to the raw viscosity data $\eta(T)$ instead of the extracted $E(T)$ or $\eta_0(T)$. This was done since  $E(T)$ and $\eta_0(T)$ are approximate measures derived from the raw data. In Fig. \ref{fig: eeta0part}, we provided extracted finite difference numerical values of $E(T)$ and $\eta_0(T)$ for six selected liquids above their liquidus temperature $T_l$. In this Appendix, we separately plot $E(T)$ and $\eta_0(T)$ of all tested metallic liquids above their liquidus temperature $T_l$ and scaling temperature $T_{sc}$ in Figs. \ref{fig:e_eta0_s_metallic} and \ref{fig:e_eta0_s}.

In Table \ref{tab:22}, we provide statistics regarding the quality of our fits. These figures of merit, computed for both the raw and filtered data, are comprised of R-square ($R^2$) values and the sum of squared errors (SSE).

\section{Further comments on estimating $E$ and $\eta_{0}$ values from fits}
\label{estimateEeta}

As discussed in the main text, \myref{eq:2} enables the extraction of the effective $E$ and $\ln\eta_0$ at different temperatures. Our filtered data were chosen to be equally spaced allowing us to fit adjacent data points with \myref{eq:2}. For example, for the temperature domain $[T_i, T_{i+1}]$ and the corresponding range of viscosity $[\ln\eta_i, \ln\eta_{i+1}]$, the fitting outputs will be $E_i$ and $\ln\eta_{0,i}$ at the temperature $T_{i+1/2}$. Similarly, for the temperature interval $[T_{i+1}, T_{i+1+1}]$, the outputs are $E_{i+1}$ and $\ln\eta_{0,i+1}$ at $T_{i+1+1/2}$, etc.

Our initial analysis centered on temperature intervals of width $\Delta T=T_{i+1}-T_i=0.25K$. Such relatively small temperature intervals $\Delta T$ increase the uncertainties in $E$ and $\ln\eta_0$ and introduce oscillations in $E(T)$ and $\ln\eta_0(T)$. To avoid these spurious effects and generate more monotonous trends, we varied the width of the temperature intervals over which we compute the averages. As this latter width increases, the resulting $E(T)$ and $\ln\eta_0(T)$ curves become smoother as more fluctuations are removed. When this factor is large enough, the general monotonic trends of $E$ and $\ln\eta_0$ become lucid. For instance, if this scaling factor is $60$ then the temperature intervals will be $[T_{60i},T_{60(i+1)}]$. 

Fig. \ref{fig:zrptfir} shows, respectively, the values of $E(T)$ and $\ln\eta_0(T)$ for $\rm Zr_{80}Pt_{20}$ that were obtained in this way. The temperature window scaling factor is 60 with a corresponding the temperature interval width $\Delta T =15K$. As the temperature rises from $1400K$ to $1900K$, the effective activation energy $E$ drops from $11000K$ to $6000K$ whereas $\ln \eta_0$ increases from $-10.7$ to $-9$. As we noted earlier, from \myref{eq:3} (a trivial restatement of \myref{eq:2}), when $T\ln\eta$ is examined as a function of the temperature, the activation energy $E$ becomes the intercept and $\ln\eta_0$ is the slope.

\section{Additional Viscosity Fits}

\begin{figure}[t]
	\centering
	\includegraphics[scale=1]{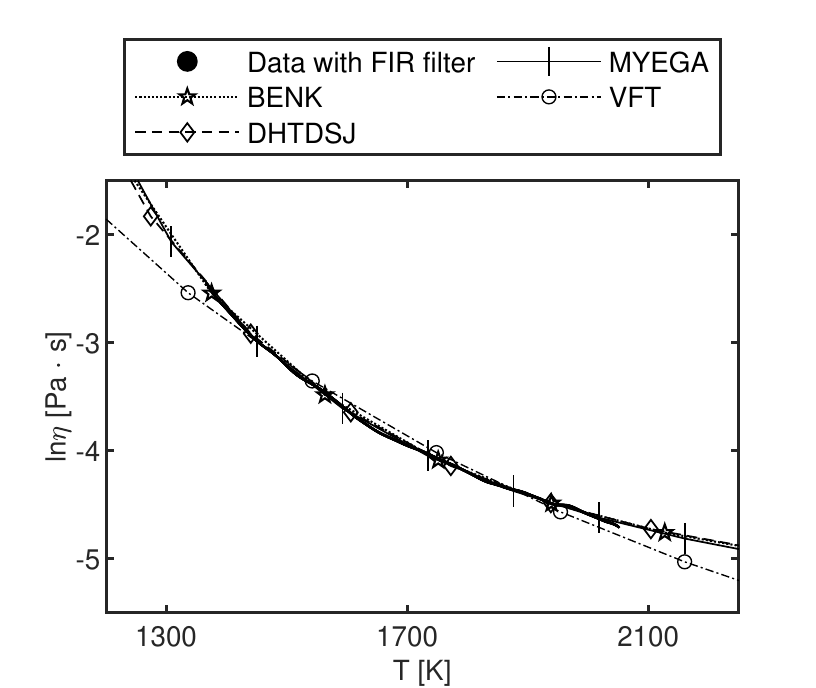}
	\caption{(Color online.)  A comparison between our experimentally measured filtered viscosity data of \ce{Zr80Pt20} with four fitting forms: BENK (\myref{con:benk}), DHTDSJ (\myref{ref:DHTDSJ}), MYEGA (\myref{ref:MYEGA}), and VFT (\myref{con:vft}).}
	\label{fig:compareothers}
\end{figure}

Aside from the MYEGA \cite{myega1} and DHTDSJ \cite{dhtdsj1} fits of Eqs. (\ref{ref:MYEGA}, \ref{ref:DHTDSJ}) (the inception of which was motivated by the behavior of glassformers), there are numerous other fitting forms that attempt to describe the viscosity of fluids at high temperatures (as well as the viscosities of bona fide supercooled liquids at temperatures below equilibrium freezing). 
For concreteness, we list several of these below. 

We start by noting perhaps by the far most common empirical form, that of Vogel, Fulcher, and Tammann (VFT) \cite{vft1,vft2,vft3}. Herein,
\begin{equation}
\ln\eta(T)=\ln\eta_0+\frac{B}{T-T_A}
\label{con:vft}
\end{equation}
with material dependent parameters $\eta_0$, $B$ and $T_A$.

According to the KKZNT \cite{kkznt1, kkznt2, kkznt3} fit for the viscosity 
\begin{equation}
\ln\eta(T)=\ln\eta_0+\frac{E_{\infty}}{T}+\frac{T_A}{T}B \Big( \frac{T_A-T}{T_A} \Big)^z\Theta(T_A-T),
\label{con:kkznt}
\end{equation}
where $\eta_0,E_{\infty}, T_A$ and $z$ are liquid dependent adjustable constants (with, in most fluids, $z \simeq 8/3$).

The DEH \cite{deh1} fit asserts that
\begin{equation}
\ln\eta(T)=\ln\eta_0+\frac{E_{\infty}}{k_BT}+\frac{(T-T_A)^2}{2a^2}\Theta(T_A-T),
\label{con:deh}
\end{equation}
with $\eta_0, E_{\infty}, a,$ and $T_A$ being material dependent parameters.

Another functional form (BENK) \cite{Tcoop1} that we studied suggests that
\begin{equation}
\ln\eta(T)=\ln\eta_0+\frac{E}{k_BT}+J^2(\frac{1}{T}-\frac{1}{\widetilde{T}})^2\Theta(T_A-T).
\label{con:benk}
\end{equation}
Here, the adjustable, fluid dependent, constants are $\eta_0, J, \widetilde{T}$ and $T_A$. 

Numerous functional forms and theoretical approaches including, in particular those related to the enigmatic glass transition, appear in the literature, e.g., \cite{Langer2014,supercool1996, supercool1,supercool2,Angell1924,relaxAngell,qmrelaxtime1,kelton2016,correlation2016,relaxArr2012, berthier2011, BerthierBiroli2011}. Some of these, similar to the above forms build on Arrehius type notions and various modifications of this form. An Arrhenius type analysis was recently pursued in \cite{kshirai2020},
\begin{equation}
\ln\eta(T)=\ln\eta_0 -\frac{b}{k_B}+\frac{Q_a^*}{k_BT}.
\label{con:ksqa}
\end{equation}
Here, $Q_a^*$ is the effective activation energy computed by \myref{eq:4}. When computed form viscosity data of supercooled liquids, this effective energy barrier exhibits a peak around the glass transition $T_{g}$. This led \cite{kshirai2020} to posit that the glass transition is associated with a bona fide phase transition at $T_g$.

In Eqs. (\ref{con:vft}, \ref{con:kkznt}, \ref{con:deh}, \ref{con:benk}), $T_A$ (see also a brief discussion in the main text) denotes a crossover temperature from a putative Arrhenius behavior ($T > T_{A}$) to super-Arrhenius scaling ($T< T_{A}$). 
\myref{con:ksqa} does not invoke a crossover temperature. 

Our results concerning the high temperature deviations from activated dynamics (including temperatures above assumed crossover temperatures) suggest that the VFT, KKZNT, DEH and BENK fits and similar others that assume high temperature Arrhenius dynamics do not accurately describe high temperature liquids. Our finding does not exclude the asserted functional forms
of these fits at all temperatures- only their behaviors at high temperatures. By contrast, fits like those of Eqs. (\ref{ref:MYEGA}, \ref{ref:DHTDSJ}) that, at all $T$ (in particular, also for all temperatures above equilibrium melting), exhibit an effective activation barrier $E(T)$ that monotonically decreases with increasing temperature are consistent with the trends that we universally find in all examined high temperature fluids.

\section{Low Pass Filter}
\label{lowpassfilter}

\begin{figure}[t]
	\centering
	\includegraphics[scale=1]{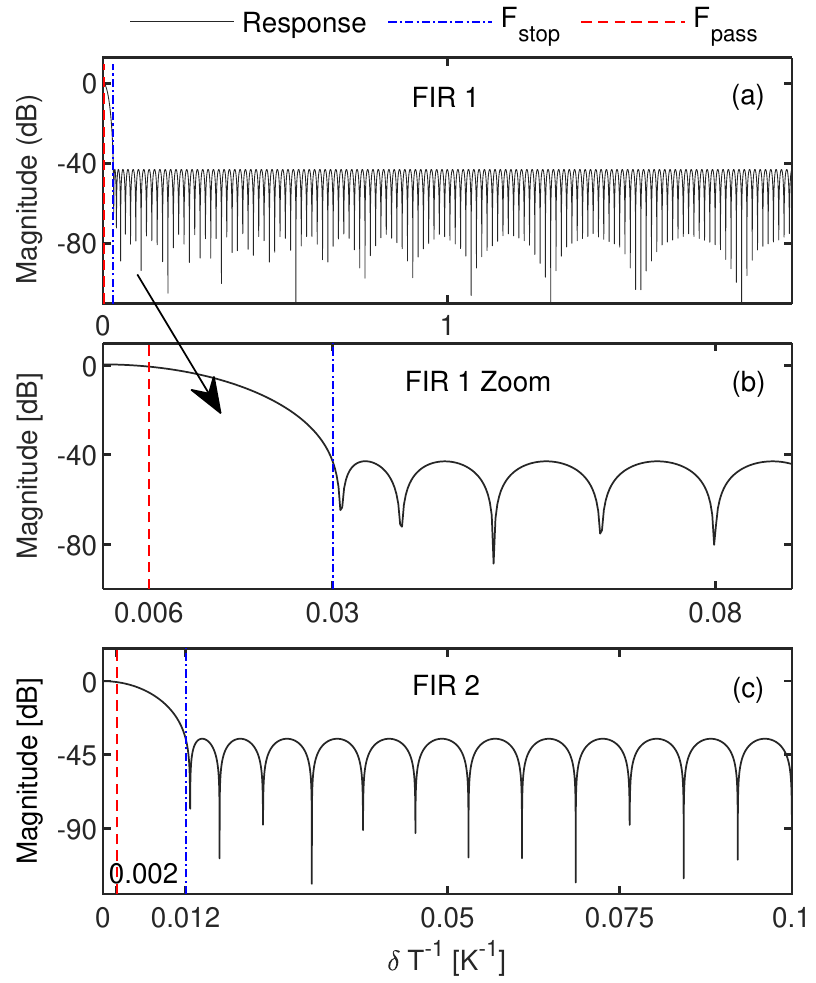}
	\caption{(Color online.)	The amplitude response of two low-pass FIR filters (labeled FIR 1 and FIR 2). The abscissa denotes the number of samples per inverse temperature interval, $1/ \delta T$. The ordinate marks the magnitude of attenuation of the filter. The red and blue dashed lines indicate, respectively, the sample rate at $F_{pass}$ and $F_{stop}$. As all panels of this figure illustrates, the (black) response curve starts to oscillate for sample rates beyond $F_{stop}$. Panel (b) is a blow up of FIR 1's amplitude response (Panel a). Here, $F_{pass}=0.006K^{-1}$ and $F_{stop}=0.03K^{-1}$. In Panel (c), we provide a blow up of FIR 2's response with $F_{pass}=0.002K^{-1}$ and $F_{stop}=0.012K^{-1}$.
	}
	\label{fig:fir}
\end{figure}

\begin{table}[!h]
	\centering
	\caption{The low pass FIR filters that we used to filter the viscosity data. FIR1 filters the raw viscosity data with $\Delta T=0.25K$. For FIR2, $\Delta T = 5K$ and $15K$. Both FIR1 and FIR2 are generated by the software MATLAB.}
	\vspace{1mm}
	\renewcommand\arraystretch{1.2}
	\tabcolsep0.14in
	\begin{tabular}{lll}
		\hline
		Filter&			FIR1&			FIR2\cr
		Specify Order&		260&			25\cr
		$F_s$&			4$K^{-1}$&		0.067$K^{-1}$\cr
		$F_{pass}$&		0.006$K^{-1}$&	0.002$K^{-1}$\cr
		$F_{stop}$&		0.03$K^{-1}$&		0.012$K^{-1}$\cr
		\hline \cr
	\end{tabular}
	\label{tab:fir33}
\end{table}

In the current work, we invoked ideas similar to those used in standard frequency filters and (given that our viscosity data are a function of temperature and not time) extended these to the temperature domain (i.e., with $\frac{1}{\Delta T}$ playing the role of frequency in the typically used filters \cite{ingle1997digital}). Figs. \ref{fig:fir} and Table. \ref{tab:fir33} provide basic schematics of our low pass FIR filter (FIR1) for the viscosity data, while Figs. \ref{fig:fir} provide the schematics of our second FIR filter for determining the activation energy and effective entropy. When applying an equidistant interpolation to our viscosity data, the temperature interval between each two adjacent data points is $\Delta T=0.25K$. 
Thus, the sampling rate of the filter is $F_s=1/\Delta T=1/0.25K=4K^{-1}$. 
We then increased the width of the temperature windows over which we compute the averages and fitted the decimated data with \myref{eq:4} to obtain $E_{slope}$ and applied our second FIR filter (FIR 2). Herein, $\Delta T = 15K$, and the sampling rate of the second filter is $F_s=1/\Delta T=1/15K=0.067K^{-1}$. We repeated this procedure for $S_{slope}$ with our second FIR filter.

Figs. \ref{fig:fir} illustrate the effects of the two filters. In these figures, the abscissa is the temperature frequency $F_t=1/\delta T$ (where $\delta T$ is a temperature interval not smaller than $\Delta T$). The vertical axis is the attenuation magnitude of the filter. The minimal temperature interval used was of width $\Delta T=0.25K$ (for FIR1) and $\Delta T = 15K$ for (FIR2) and $F_{t,max}=F_s$. 

Whenever $F_t$ is smaller than $F_{pass} = 0.006K^{-1}$ (FIR1) and $0.002K^{-1}$ (for FIR2), filtering leads to no change. By contrast, when $F_{pass}<F_t<F_{stop} = 0.03K^{-1}$ (for FIR1) and $0.012K^{-1}$ (for FIR2), as $F_t$ increases, the magnitude of the filtered data monotonically decreases (with the filtered data being reduced by $42.9dB$ for FIR1 (and $35.2dB$ for FIR 2) just above $F_{stop}$. For $F_t \ge F_{stop}$, the magnitude of the filtered data remains, approximately, constant. The filter attenuates data sufficiently close to $T_i$ while leaving data far from $T_i$ essentially unchanged. Applying a low-pass FIR filter to the data leads to a smoother and more monotonous result while, concomitantly, preserving the original trends present in the data.\\

\bibliography{reff}

\end{document}